\documentclass[final,3p,times,fleqn]{elsarticle}
\usepackage{mathtools,booktabs}
\usepackage{bm}
\usepackage{autobreak}
\usepackage{amsmath}
\usepackage{amsthm}
\usepackage{siunitx}
\usepackage{gensymb}
\usepackage{subfigure}
\usepackage{epstopdf}
\usepackage{color}
\usepackage{hyperref}
\usepackage{cases}

\biboptions{sort&compress}
\begin{document}
\begin{frontmatter}
\title{Improved phase field model for two-phase incompressible flows: Sharp interface limit, universal mobility and surface tension calculation}
\author[1]{Jing-Wei Chen}
\author[1]{Chun-Yu Zhang\corref{cor1}}\cortext[cor1]{Corresponding authors.}\ead{cyzhangcfd@ustc.edu.cn}  
\author[1]{Hao-Ran Liu}
\author[1]{Hang Ding\corref{cor1}}\ead{hding@ustc.edu.cn}  
\address[1]{Department of Modern Mechanics, University of Science and Technology of China, Hefei 230027, China}
\cortext[cor1]{Corresponding author.}
\begin{abstract}
In this paper, we propose an improved phase field model for interface capturing in simulating two-phase incompressible flows. The model incorporates a second-order diffusion term, which utilizes a nonlinear coefficient to assess the degree of deviation of interface profile from its equilibrium state. In particular, we analyze the scale of the mobility in the model, to ensure that the model asymptotically approaches the sharp interface limit as the interface thickness approaches zero. For accurate calculations of surface tension, we introduce a generalized form of smoothed Dirac delta functions that can adjust the thickness of the tension layer, while strictly maintaining that its integral equals one, even when the interface profile is not in equilibrium. Furthermore, we theoretically demonstrate that the ‘spontaneous shrinkage’ of under-resolved interface structures encountered in the Cahn-Hilliard phase field method does not occur in the improved phase field model. Through various numerical experiments, we determine the range of the optimal mobility, confirm the theoretical analysis of the improved phase field model, verify its convergence, and examine the performance of different surface tension models. The numerical experiments include Rayleigh-Taylor instability, axisymmetric rising bubbles, droplet migration due to the Marangoni effect, partial coalescence of a droplet into a pool, and deformation of three-dimensional droplet in shear flow. In all these cases, numerical results are validated against experimental data and/or theoretical predictions. Moreover, the recommended range of dimensionless mobility ($1/Pe\sim 100 Cn^2-333 Cn^2$) has been shown to be universal, as it can be effectively applied to the simulations of a wide range of two-phase flows and exhibits excellent performance.
	\end{abstract}
\end{frontmatter}

\section{Introduction}
Accurately tracking the evolution of interfaces is essential when simulating multiphase flows. Interface capturing methods, including the level set (LS), volume of fluid (VOF), and phase field (PF) methods, utilize a scalar function known as the order parameter to define the position and shape of interfaces, which are represented by particular contours of this parameter. These methods are preferred for the simple numerical implementation and the ability to handle changes in interface topology without requiring special treatment. However, the advection of the order parameter is directly associated with local flow velocities, which are generally non-uniform along the normal direction to the interface. This inevitably causes the order parameter to deviate from its equilibrium distribution (e.g. the signed distance function in the LS methods and the hyperbolic tangent function in the PF methods), especially for long time computation. In order to accurately calculate important interface information such as normal direction and curvature, it is necessary to restore the order parameter to its equilibrium state.

Various numerical techniques have been proposed to maintain the order parameter close to its equilibrium state. For instance, re-initialization process~\cite{Sussman1994jcp, Hartmann2008jcp, Sussman1997jfm, Sethian2003arfm} is employed in the LS methods after the order parameter is advected by the flow velocity. The re-initialization process can effectively restore the order parameter to the signed distance function. However, it should be noted that the re-initialization tends to bring in additional numerical errors such as mass loss~\cite{Sussman1994jcp, Luo2015jcp}. Although there is no separate re-initialization process in the PF methods, the diffusion term in the Cahn-Hilliard equation, which originates from the microscopic theories of fluid-fluid interfaces, plays a role similar to its effect. The diffusion term can counteract the distortion effects to the interface profile owing to the non-uniform flow velocity, and at the same time, maintains the volume conservation of each phase~\cite{Jacqmin1999jcp, Ding2007jcp, Kim2012ccp}. Despite the successes of the LS and PF methods, they share one challenge in common: how to properly reinitialize the order parameter~\cite{Russo2000jcp, Magaletti2013jfm}. Inadequate reinitialization fails to recover the order parameter to the equilibrium state, while excessive reinitialization may result in significant mass loss or unrealistic interface movement.

Recently, the improved PF method has become increasingly favored for simulating incompressible multiphase flows~\cite{Olsson2005jcp, Olsson2007jcp, Chiodi2017jcp, Janodet2022jcp, Doherty2023jcp, Wang2023jcp, Abadi2018jcp, Jettestuen2021jcp}. The method was proposed by Olsson and Kreiss~\cite{Olsson2005jcp} to enhance mass conservation in the LS method. Since the re-initialization process in the improved PF method incorporates a conservative diffusion term (and is mathematically equivalent to the solution of Burgers equation~\cite{Olsson2007jcp}), it can ensure the conservation of the volume bounded by the diffuse interface. The method was initially termed the conservative LS method, as the advection of the order parameter and the re-initialization process are treated separately, as is done in the LS method. Later, the two processes were combined~\cite{Chiu2011jcp} by decomposing the normal interface speed into two parts that are independent of (due to external advection) and proportional to the interface curvature. In contrast to the fourth-order derivative in the PF method~\cite{Jacqmin1999jcp}, the diffusion term in the improved PF method only includes second-order derivatives, thereby significantly reducing the requirement for the time step to maintain computational stability. However, how to determine the appropriate value of the mobility $M$, which measures the relative significance of the diffusion term to the flow advection, is still under intensive research. Both excessively small and large amounts of numerical diffusion can introduce unwanted interface problems, similar to those encountered in the LS and PF methods.  Chiu and Lin~\cite{Chiu2011jcp} recommended to choose the maximum dimensionless speed of the flow for the mobility, i.e. $M\sim|\widetilde u_{max}|$. Nevertheless, it has been shown that the mobility must be adjusted on a case-by-case basis~\cite{Chiu2011jcp, Chiu2019jcp}. Furthermore, to achieve the boundedness of the order parameter,  Mirjalili et al.~\cite{Mirjalili2020jcp} proposed a criterion between $M$ and the interface thickness $\epsilon$: $\epsilon/h\ge(M/|\widetilde u_{max}|+1)/(2M/|\widetilde u_{max}|)$, where $h$ is the grid size. To the best of our knowledge, establishing a universal rule for the mobility $M$ remains an open question. 

It is known that surface tension plays an important role in two-phase incompressible flows, in particular for flows driven by surface tension such as thermocapillary migration~\cite{Young1959jfm, Karbalaei2016micromachines, Xu2021jfm, Zeng2021pans, Hu2022nc} and flows with moving contact lines~\cite{Clanet2004jfm, Ding2012jfm, Sui2014arfm, Snoeijer2013arfm, Ding2018jfm, Zhang2023jfm}. The continuous surface force (CSF) model is often employed in diffuse interface simulations of two-phase flows~\cite{Brackbill1992jcp, Popinet2018arfm, Yang2021jcp, Jabir2021jcp, Jain2022jcp, Mirjalili2024jcp}. This model essentially transforms the surface tension at fluid-fluid interfaces into the integration of a corresponding volumetric force by integrating a smoothed Dirac delta function~\citep{Brackbill1992jcp}, which is a function of the order parameter. Consequently, the precision in calculating surface tension is dependent not only on the accurate computation of the interface’s normal direction and curvature but also on a suitable $\delta$ function, especially when the interface profile is out of equilibrium. In principle, to replicate the stress jump condition in the sharp interface limit, the $\delta$ function must adhere to the fundamental property of the Dirac delta function, i.e. the integral from $-\infty$ to $+\infty$ along the normal direction to the interface must equal one, regardless of the distribution of the interface profile. In the improved PF method, because the distribution of the volume fraction $C$ at equilibrium is a hyperbolic tangent function, the differentiation of $C$ naturally acts as a smoothed Dirac delta function, e.g. $\delta=|\nabla C|$~\cite{Olsson2005jcp, Yang2021jcp, Brackbill1992jcp}, in which 90\% of the surface tension is within the interface region (e.g. $0.05 \leq C \leq 0.95$). To reduce the thickness of the tension layer, another smoothed delta function $\delta=6\sqrt{2}\epsilon |\nabla C|^2$ was proposed~\cite{Kim2005jcp, Liu2013pre}, which accounts for about 98.5\% of the surface tension force in the interface region. However, this $\delta$ function fails to uphold the fundamental property of the Dirac delta function when the interface profile deviates from the equilibrium state (see the details in Section~\ref{Surface tension model}). Based on the above analysis, it is recommended to use Dirac delta functions that strictly conform to the conditions of the original Dirac delta function and have adjustable thickness, although such functions are not currently available in the literature. 

In this paper, we first derive an improved PF model that incorporates a second-order diffusion term, with a diffusion coefficient measuring the deviation of the interface profile from the equilibrium state. We then evaluate the scale of mobility to ensure that the model asymptotically approaches the sharp interface limit as the interface thickness approaches zero. Furthermore, we propose a general form of $\delta$ functions related to the $C$ field, which can be adjusted in thickness while strictly ensuring that its integral equals one. Also, we theoretically analyze the performance of the improved PF method in the presence of under-resolved interfaces, such as tiny droplets and bubbles, demonstrating that the ‘spontaneous shrinkage’ observed in the Cahn-Hilliard PF method does not occur in the improved PF model. Combining the improved PF model with the incompressible Navier-Stokes equations, we can simulate two-phase flows with large density ratios. Through various numerical experiments, we determine the range of the optimal mobility, confirm the theoretical analysis of the improved PF model, verify its convergence, and examine the performance of the surface tension models. The numerical experiments include Rayleigh-Taylor instability, axisymmetric rising bubbles, droplet migration due to the Marangoni effect, partial coalescence of a droplet into a pool, and deformation and breakup of three-dimensional droplets in a simple shear flow. 

\section{Interface models}
\label{governing}
\subsection{Cahn-Hilliard phase field model}
In the PF models, the mathematically sharp interface separating two immiscible fluids is replaced with a diffuse interface with finite thickness (measured by a prescribed parameter $\epsilon$). In particular, in the Cahn-Hilliard (CH) PF model, the continuous variations of the order parameter, which represents the interface, are consistent with the gradient theories of the interface based on thermodynamic principles~\cite{Van1979jsp, Cahn1959jcp}. Consequently, for incompressible two-phase flows, the temporal evolution of the order parameter, e.g. the volume fraction of one fluid $C$ $(0\le C \le 1)$, can be described by~\cite{Ding2007jcp},
\begin{equation}
\frac {\partial C}{\partial t}+\nabla\cdot\left({\bf u} C\right)=M\nabla^2\left(\psi'(C)-\epsilon^2\nabla^2C\right),
\label{CH_Ding2007}
\end{equation}
where $M$ is the mobility, $\phi=\psi'(C)-\epsilon^2\nabla^2C$ the chemical potential, and $\psi(C)=C^2(1-C)^2/4$ the bulk energy density~\cite{Jacqmin1999jcp, Ding2007jcp}. For an interface at equilibrium, the chemical potential is uniformly distributed in the domain ($\phi=0$). Moreover, the interface profile at equilibrium for a flat interface can be analytically obtained,
\begin{equation}
C_{eq}(z)=\frac{1}{2}+\frac{1}{2}\tanh\left(\frac{z-z_0}{2\sqrt{2}\epsilon}\right),
\label{phase_function}
\end{equation}
where $z$ is the coordinate along the gradient of $C$, and $z_0$ corresponds to the position of the contour $C=0.5$. In order to asymptotically approach the sharp interface limit, in principle the right term in Eq.~(\ref{CH_Ding2007}) should vanish when the interface thickness goes to zero. This can be achieved by adjusting the mobility according to the value of $\epsilon$, e.g. $M\propto\epsilon^2$~\cite{Magaletti2013jfm}.

Despite the fact that the CH model originates from the microscopic theories of the interface, the term on the RHS of Eq.~(\ref{CH_Ding2007}) serves as 'effective reinitialization' that restores the distribution of the $C$ field to the equilibrium state from the computational point of view. Furthermore, the value of chemical potential $\phi$ provides an effective evaluation of the deviation of the interface profile from the equilibrium state. For example, a negative $\phi$ suggests a stretched interface while a positive one corresponds to a compressed interface. However, there are several noticeable disadvantages of the CH model. Firstly, under the conditions of significant velocity gradients across the interface (e.g. rising bubbles due to buoyancy), it is difficult for the term on the RHS of Eq.~(\ref{CH_Ding2007}) to restore the interface to the equilibrium state without artificially changing the interface shape. Similarly, simulation of flows with Marangoni effect, which strictly requires the interface profile at equilibrium, turns out to be a tough job for the CH model. Secondly, the CH model suffers from spontaneous shrinkage~\cite{Yue2007jcp}, i.e. a droplet/bubble apparently dissolves into the bulk of the other phase (bulk diffusion) when the interface thickness is of the same order as its radius. In this case, there is no equilibrium state available for the interface. The phenomenon of the bulk diffusion can be understood for a circular droplet with radius $R_0$ in the cylindrical coordinate ($r, \theta$). Assuming that the $C$ field for the droplet is uniformly distributed in the circumferential direction, i.e. $\partial^2C/\partial \theta^2=0$, the equilibrium condition $\phi=0$ gives
\begin{equation}
\frac{\partial^2C}{\partial r^2}+\frac{1}{r}\frac{\partial C}{\partial r}+\frac{\psi'(C)}{\epsilon^2}=0.
\label{PDE01}
\end{equation}
The solution of Eq.~(\ref{PDE01}) can be approximately written as the equilibrium state, i.e. Eq.~(\ref{phase_function}), only when $\epsilon$ is much smaller than the droplet radius $R_0$ so that $r^{-1}\partial C/\partial r$ is negligible. Clearly, the solution gradually deviates from the equilibrium state with increasing $\epsilon/R_0$; moreover, a solution similar to Eq.~(\ref{phase_function}) even does not exist when $\epsilon$ is approaching $R_0$, thereby leading to the occurrence of the spontaneous shrinkage. Thirdly, the presence of the fourth-order diffusion term in the CH model imposes a constraint on time step in the computation, and thus requires the use of an implicit scheme to improve the computational efficiency~\cite{Furihata2001NumMath, Liu2021jcp}.

\subsection{Improved phase field model}
Inspired by the spirit of CH model, we propose an improved PF model, in which the term on the RHS of the CH model is replaced by a second-order diffusion term,
\begin{equation}
\frac {\partial C} {\partial t}+\nabla \cdot ({\bf u}C)=M^*\nabla \cdot\left(\xi\nabla C\right),
\label{modify_PFM}
\end{equation}
where $M^*$ is the mobility in the present improved PH model, and $\xi$ is a diffusion coefficient measuring the deviation of the interface profile from the equilibrium state. To select an appropriate form of $\xi$, several principles should be followed. Firstly, $\xi=0$ if the interface profile reaches the equilibrium state. Secondly, the sign of $\xi$ is directly associated with the state of the interface profile; specifically, $\xi>0$ (diffusion) for the compressed interface and $\xi<0$ (anti-diffusion) for the stretched interface. Thirdly, the magnitude of $\xi$ should reflect the degree to which the interface deviates from the equilibrium. Moreover, given an interface at equilibrium as in Eq.~(\ref{phase_function}), it is straightforward to obtain 
\begin{equation}
|\nabla C|=\frac{\partial C}{\partial z}=\frac{C (1-C)}{\sqrt{2}\epsilon}.
\label{gradient_phi}
\end{equation}

Therefore, we can define $\xi$ as
\begin{equation}
\xi=1-\frac{C (1-C)}{\sqrt{2}\epsilon\,|\nabla C|},
\label{xi_sim2}
\end{equation}
where $|\nabla C|$ is evaluated numerically, and $\xi=0$ for either $C=1$ or $C=0$. Substituting Eq.~(\ref{xi_sim2}) into Eq.~(\ref{modify_PFM}), we get an improved PF model for incompressible two-phase flows
\begin{equation}
\frac {\partial C} {\partial t}+\nabla \cdot ({\bf u}C)=M^* \nabla \cdot\left(\nabla C-\frac{C(1-C)}{\sqrt{2}\epsilon}{\bf n}\right),
\label{ipf}
\end{equation}
where ${\bf n}=\nabla C/|\nabla C|$ is the unit vector normal to the interface. Because the RHS of Eq.~(\ref{ipf}) is a second-order diffusion term, the improved PF model can quickly restore the interface profile to the equilibrium state without distorting the physical interface shape. In addition, the spontaneous shrinkage would not occur in the improved PF model, owing to the fact that the issue of $r^{-1}\partial C/\partial r$ does not exist in the multi-dimensional problems for the interface at equilibrium (i.e. $\xi=0$). Moreover, it can be expected that the interface profile still follows Eq.~(\ref{phase_function}), even when the interface is under-resolved.

\subsection{Sharp interface limit}
Similar to the role of the mobility $M$ in the CH model, the value of $M^*$ in the improved PF model is chosen in a manner to ensure the sharp interface limit. In the sharp interface limit, all the flow features such as the interface curvature are well resolved by the interface thickness, and the term on the RHS of Eq.~(\ref{modify_PFM}) should vanish with $\epsilon \rightarrow 0$. Therefore, we need to analyze the scale of the term $\nabla \cdot\left(\xi\nabla C\right)$, in order to provide an appropriate evaluation of $M^*$. It is clear that the coefficient $\xi$ has a scale of $O(1)$ from Eq.~(\ref{xi_sim2}), and that $|\nabla C|$ has a scale of $O(\epsilon^{-1})$ from Eq.~(\ref{gradient_phi}). Given the interface curvature $\kappa=\nabla \cdot {\bf n}=\nabla \cdot (\nabla C/|\nabla C|)$, we can obtain $\nabla \cdot\left(\xi\nabla C\right)\sim \epsilon^{-1} \kappa$. It is reasonable to assume that interface curvature has a finite value in numerical simulations; if there is any singularity at the interfaces such as topology change, it only occurs in a very short duration. Therefore, if the PF model can approach the sharp interface limit, the mobility $M^*$ is expected to at least have a scale of
\begin{equation}
M^*\sim \lambda \epsilon^2,
\label{M_sil}
\end{equation}
where $\lambda$ is a constant with the dimension of the inverse of time. Accordingly, we can easily get $M^*\epsilon^{-1}\kappa\rightarrow 0$ when $\epsilon \rightarrow 0$.

\section{Governing equations and surface tension calculation}
\subsection{Governing equations}

For incompressible two-phase flows, we define the dimensionless numbers with the properties of fluid 1, characteristic length $L$ and characteristic velocity $U$: Reynolds number $Re =\rho_1 UL/\mu_1$, Froude number $Fr = U^2/(gL)$ and Weber number $We = \rho_1 U^2 L/\sigma$, where $\sigma$ denotes the surface tension coeﬀicient at the interface between the fluids 1 and 2. We also refer to the Ohnesorge number $Oh = \sqrt{We}/Re = \mu_1/\sqrt{\rho_1 \sigma L}$ and capillary number $Ca = We/Re = \mu_1 U/\sigma$. Same as the CH-PF model~\cite{Ding2007jcp}, the dimensionless governing equations for incompressible two-phase flows with large density ratio are written as
\begin{equation}
\rho\left(\frac{\partial {\bf u}}{\partial t}+{\bf u} \cdot \nabla {\bf u}\right) = -\nabla p +\frac{1}{Re}\nabla \cdot \left[ \mu \left(\nabla {\bf u}+\nabla {\bf u}^T\right)\right]+\frac{1}{We}{\bf f}_s+\frac{1}{Fr}\rho{\bf g},
\label{momentum}
\end{equation}
\begin{equation}
\nabla \cdot {\bf u} = 0,
\label{mass}
\end{equation}
where $\bf{g}$ is gravitational acceleration, and ${\bf f}_s$ is surface tension force, of which the detailed calculation is described in Section~\ref{Surface tension model}. The dimensionless average density $\rho$ and viscosity $\mu$ are defined as, 
\begin{equation}
\rho = C + (1-C)\frac{\rho_2}{\rho_1},
\label{rho dimensionless}
\end{equation} 
\begin{equation}
\mu =  C + (1-C)\frac{\mu_2}{\mu_1},
\label{mu dimensionless}
\end{equation} 
where the subscripts 1 and 2 denote the properties of respective fluids.

The dimensionless improved PF model in Eq.~\ref{ipf} for incompressible two-phase flows can be expressed as
\begin{equation}
\frac {\partial C} {\partial t}+\nabla \cdot (C {\bf u})=\frac{1}{Pe}\nabla \cdot \left( \nabla C - \frac{C (1-C)}{\sqrt{2}Cn}{\bf n}\right),
\label{mpf}
\end{equation}
where the Cahn number $Cn$ ($=\epsilon/L$) is the dimensionless measure of interface thickness, and the P\'{e}clet number $Pe$ ($=UL/M^*$) represents the ratio of the magnitudes of convection and dissipation of the PF model. It is straightforward to obtain $Pe=U/(\lambda L) Cn^{-2}$ from the mobility in the sharp interface limit, i.e. Eq.~(\ref{M_sil}). However, the appropriate value of $\lambda$ remains to be determined, e.g. by checking its effect on typical interface flows.

\subsection{Surface tension model}
\label{Surface tension model}
The continuous surface force model~\citep{Brackbill1992jcp} is adopted in the calculation of surface tension, which replaces the localized interfacial force with a body force ${\bf f}_s$ in the momentum equation
\begin{equation}
{\bf f}_S=(-\sigma \kappa {\bf n}+\nabla_s \sigma)\delta,
\label{surface_tension1}
\end{equation} 
where $\nabla_s (={\bf I}-{\bf nn}\cdot \nabla)$ represents the gradient along the tangential direction of the interface, and $\delta$ denotes a smoothed Dirac delta function. To ensure the consistency in calculating the surface tension force, the integration of $\delta$ along the normal direction to the interface (denoted by $z$) should satisfy:
\begin{equation}
\int_{-\infty}^{\infty}\delta(z)~dz =1.
\label{Dirac}
\end{equation} 

\begin{figure}[!t]
\centering
\includegraphics[width=7cm]{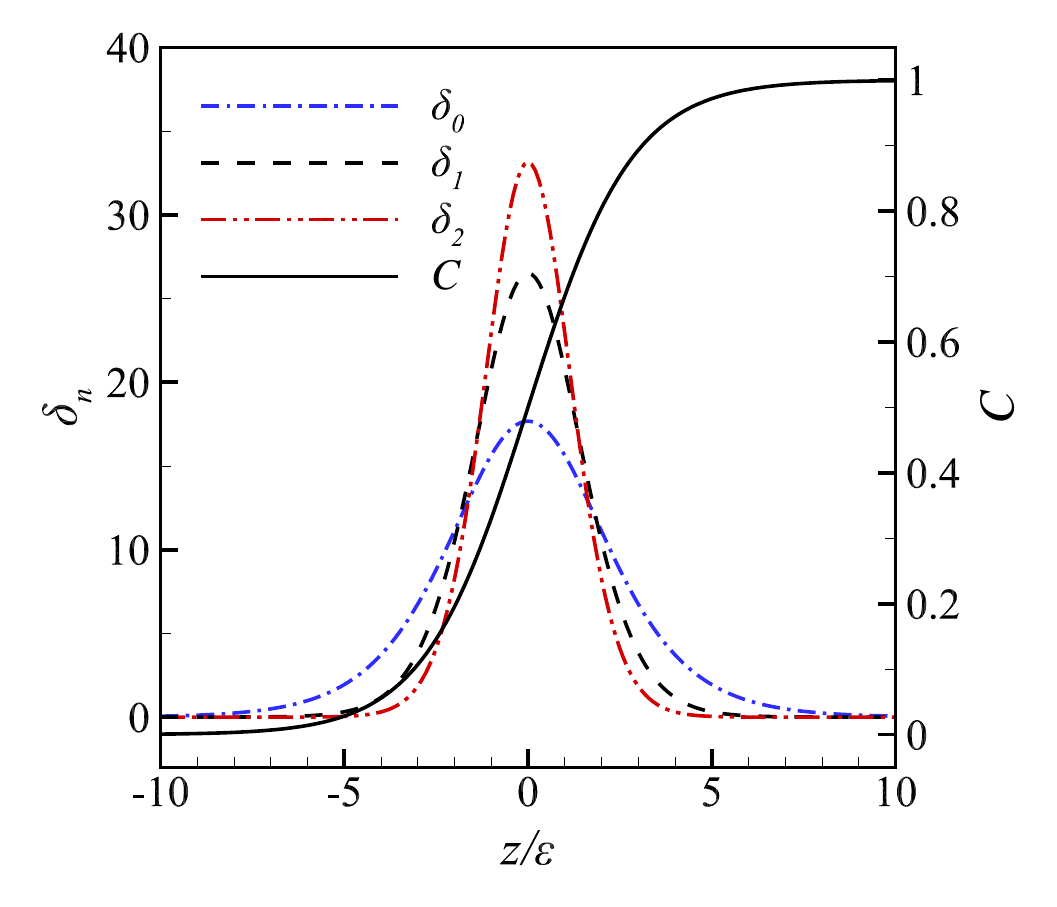}
\caption{Distribution of the smoothed $\delta_n$ functions for an interface located at $z=0$ with $\epsilon=0.01$, where $n=0$, $1$ and $2$, respectively. }
\label{dis_delta}
\end{figure}

The smoothed Dirac delta function can be directly constructed from the $C$ field. For example, it was proposed to use $\delta^{*}=6\sqrt{2}\epsilon |\nabla C|^2$ in previous studies~\cite{Kim2005jcp, Liu2013pre}, which obeys the constraint in Eq.~(\ref{Dirac}) only for interfaces in equilibrium. In order to satisfy the constraint of smoothed Dirac delta function regardless of the distribution of the interface profile, we propose a general form of $\delta_n$ with respect to the $C$ field as follows
\begin{equation}
    \delta_n= K C^n (1-C)^n |\nabla C|,
    \label{delta_n}
\end{equation}
where $K=2^{2n+1}\Gamma(3/2+n)/(\sqrt{\pi}\Gamma(n+1))$ and $\Gamma(n)$ is the gamma function. Specifically, the three low-order delta functions (i.e. $n=0 \sim 2$) yield,
\begin{equation}
\begin{cases}
\delta_0= |\nabla C|,\\
\delta_1= 6C (1-C)|\nabla C|,\\
\delta_2= 30 C^2 (1-C)^2 |\nabla C|.
\end{cases}
\label{delta}
\end{equation}
The distribution of the three delta functions for an equilibrium interface (with the $C$ profile in Eq.~(\ref{phase_function})) is shown in Fig.~\ref{dis_delta}. Clearly, the higher the number $n$, the more concentrated the surface tension. More precisely, integrating the function $\delta_n$ ($n=0$, $1$, and $2$) from $C=0.05$ to $0.95$, we can get 90.00\%, 98.55\% and 99.77\%, respectively.

It is easily demonstrated that $\delta_n$ strictly follows the constraint in Eq.~(\ref{Dirac}). Although $\delta^{*}$ in~\cite{Kim2005jcp, Liu2013pre} is equivalent to $\delta_1$ for interfaces in equilibrium, the integral value of $\int_{-\infty}^{\infty} \delta^{*} dz$ heavily depends on the distribution of the $C$ field. For example, given 1D interface profile out of equilibrium such as $C=0.5+0.5\tanh(\mathcal{R} z/(2\sqrt{2}\epsilon))$, we can obtain $\int_{-\infty}^{\infty} \delta_1 dz =1$ while $\int_{-\infty}^{\infty} \delta^{*} dz=\mathcal{R}$, where $\mathcal{R} > 1$ for compressed interface profiles and $\mathcal{R} < 1$ for stretched ones. Clearly, the use of non-consistent $\delta$ function can lead to significant numerical errors in calculating surface tension, due to the fact that the interface profile is normally maintained to be close to the equilibrium state during the computation.

\section{Numerical discretization}
\label{discret}
The discretization of the governing equations, i.e. Eqs.~(\ref{momentum}), (\ref{mass}) and (\ref{mpf}), is accomplished on a staggered grid with a uniform mesh size $\Delta x$. The scalar quantities such as pressure and volume fraction are defined at the cell centers, while the velocity components are defined at the cell faces. The mesh size is made dimensionless by $h=\Delta x/L$. 

A third-order Runge-Kutta method~\cite{Shu1988jcp} is used for temporal discretization of the improved PF equation, Eq.~(\ref{mpf}). In each step, the discretization can be written as 
\begin{equation}
\frac{C ^{n+1}-C ^{n}}{\Delta t}=-\nabla \cdot (C^n {\bf u}^n)+\frac{1}{Pe}\nabla\cdot F^n,
\end{equation}
where ${\bf F}= \nabla C - C (1-C)/(\sqrt{2}Cn){\bf n}$ denote the diffusion flux. The finite volume method is implemented for the spatial discretization, and thus the fluxes at the cell faces need to be approximated. Taking the approximation of the diffusion flux in two-dimensional form, ${\bf F}_{i+1/2, j}$, as an example, 
\begin{equation}
    F_{i+1/2,j} \approx \frac{C_{i+1,j}-C_{i,j}}{h}-\frac{S^x_{i+1,j}+S^x_{i,j}}{2} + O(h^2),
\end{equation}
where the vector ${\bf S}=C (1-C)/(\sqrt{2}Cn){\bf n}$ is defined at the cell center, and $S^x$ denotes its $x$ component. The term $\nabla C$ in ${\bf S}$ is approximated by a second-order scheme; more specifically, the discretization of its $x$ component, $\partial C/\partial x$, can be expressed as
\begin{equation}
\left(\frac{\partial C}{\partial x}\right)_{i, j}\approx\frac{C_{i+1, j+1}+2 C_{i+1, j}+C_{i+1, j-1}-C_{i-1, j+1}-2 C_{i-1, j}-C_{i-1, j-1}}{4h}+O(h^2).
\end{equation}
The value of $C$ in the convection term $\nabla \cdot (C {\bf u})$ is also required at the cell face, and is obtained by using a 5th-order WENO scheme~\cite{Liu1994jcp}.

The coupling of the momentum equation and the continuity equation is performed by a standard projection method. First, the momentum equation is solved using the Adams-Bashforth scheme for the convection term and the Crank-Nicolson scheme for the diffusion term, to obtain an intermediate velocity ${\bf u}^*$:
\begin{equation}
\frac{{\bf u}^* - {\bf u}^n}{\Delta t} = \frac{1}{\rho^{n+1/2}} \left\{ - \left[ \frac{3}{2} H\left({\bf u}^n\right) - \frac{1}{2} H\left({\bf u}^{n-1}\right) \right] + \frac{1}{2 Re} \left[ D\left({\bf u}^*, \mu^{n+1}\right) + D\left({\bf u}^n, \mu^{n}\right) \right] \right\},
\label{discretization of NS_1}
\end{equation}
where $H$ denotes the discrete convection operator and $D$ the discrete diffusion operator. The intermediate velocity is corrected by
\begin{equation}
\frac{{\bf u}^{n+1} - {\bf u}^n}{\Delta t} = -\frac{\nabla p^{n+1/2}}{\rho^{n+1/2}}.
\label{discretization of NS_2}
\end{equation}
Then, the pressure can be obtained by the Poisson equation arising from the divergence-free velocity constraint, i.e.
\begin{equation}
\nabla \cdot \left( \frac{\nabla p^{n+1/2}}{\rho^{n+1/2}} \right) = \frac{\nabla \cdot {\bf u}^*}{\Delta t}.
\label{discretization of NS_3}
\end{equation}
The QUICK scheme~\cite{Leonard1979cmame} is employed for spatial discretization of the convection operator, while the central difference scheme is used for the diffusion operator. 

For the calculation of surface tension force, the Simpson Rule is adopted for the volume integration of $f$,
\begin{equation}
\int_{\Omega_{i+1/2, j}} f d\Omega \approx \frac{h^2}{6}(f_{i, j} + 4f_{i+1/2, j} + f_{i+1, j})+O(h^3),
\label{Eq:Simpson}
\end{equation}
where $\Omega$ denotes the control volume centered at $(i+1/2, j)$.

\begin{figure}[!t]
\centering
\includegraphics[width=6cm]{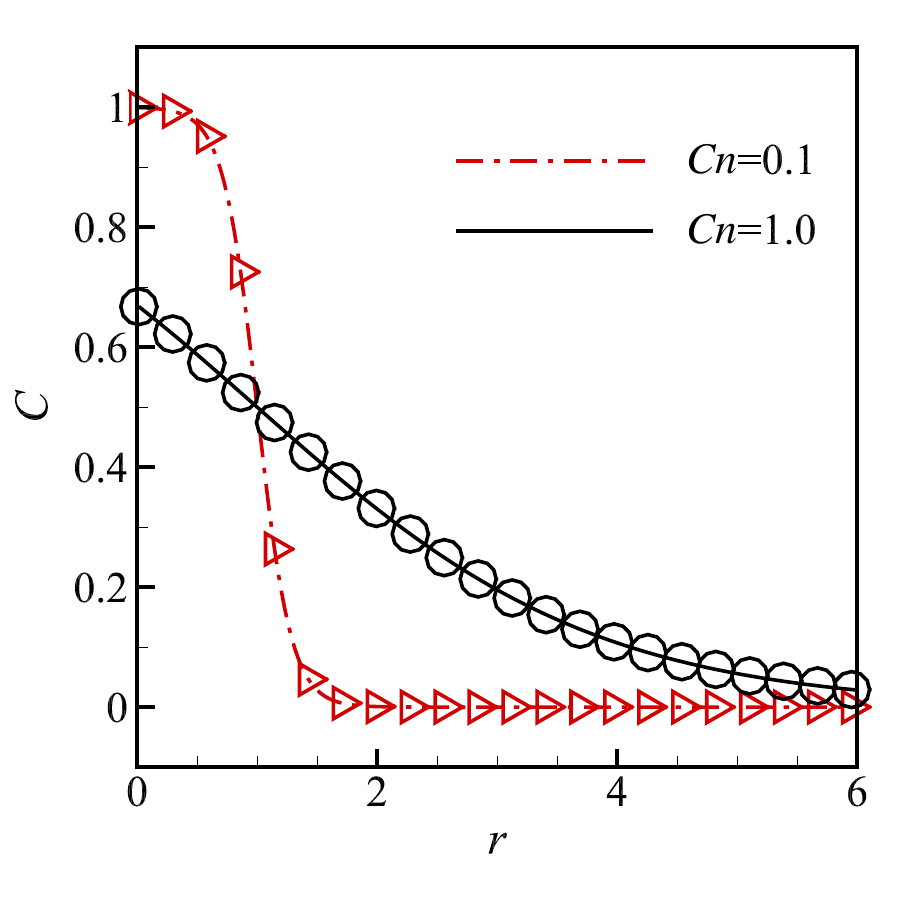}
\put(-180,160){($a$)}
\hspace{1cm}
\includegraphics[width=6cm]{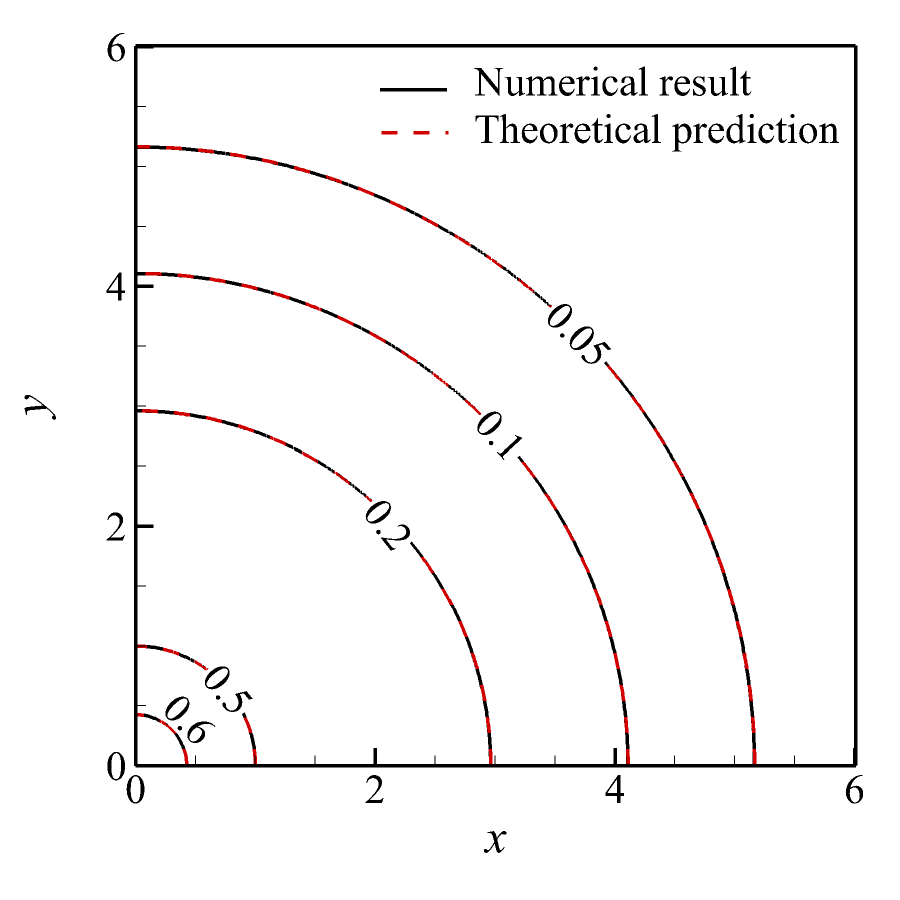}
\put(-180,160){($b$)}
\hspace{1cm}
\caption{Distribution of the $C$ field of an under-resolved droplet: the droplet radius $R=1$, the mesh size $h=0.02$, and the $Pe$ number $Pe=0.01Cn^{-2}$. ($a$) Radial distribution of $C$ with $Cn=0.1$ and $Cn=1$, where the symbols denote numerical results and the curves represent the equilibrium state. (b) Spatial distribution of $C$ with $Cn=1$, where the solid curves denote the numerical results in terms of the $C$ contours and the dashed lines are the theoretical prediction of the equilibrium state.}
\label{case01.Unsolve}
\end{figure}

\section{Results and discussion}
\label{Results and discussion}
\subsection{Under-resolved interfaces}
\label{subsection:under-resolved interfaces}
We evaluate the performance of the present method in the presence of under-resolved interfaces, of which the curvature cannot be resolved by the interface thickness. In this study, we only solve the improved PF model, and the velocity is assumed to be zero; thus the time variation of the $C$ field is driven by the diffusion flux in the improved PF model. The test case is a stationary droplet with radius of $R=1$ and located at $(0, 0)$ in a domain of $(0, 6)\times(0, 6)$; the boundary conditions are set as follows: symmetric conditions at the left and bottom boundaries, and slip conditions at the right and top boundaries. The mesh size is fixed to $h=0.02$ and the value of $Cn$ ranges from $0.1$ to $1$. If we consider the interface thickness equals the distance between the contours $C=0.05$ and $0.95$, which accounts for $98.55\%$ surface tension with $\delta_1$, the interface thickness is roughly $8.5Cn$. Therefore, the droplet is under-resolved in the simulations with $Cn>0.12$, since its radius is smaller than the interface thickness. The P\'{e}clet number is chosen as $Pe=0.01Cn^{-2}$.

\begin{figure}[!t]
\centering
\includegraphics[width=6cm]{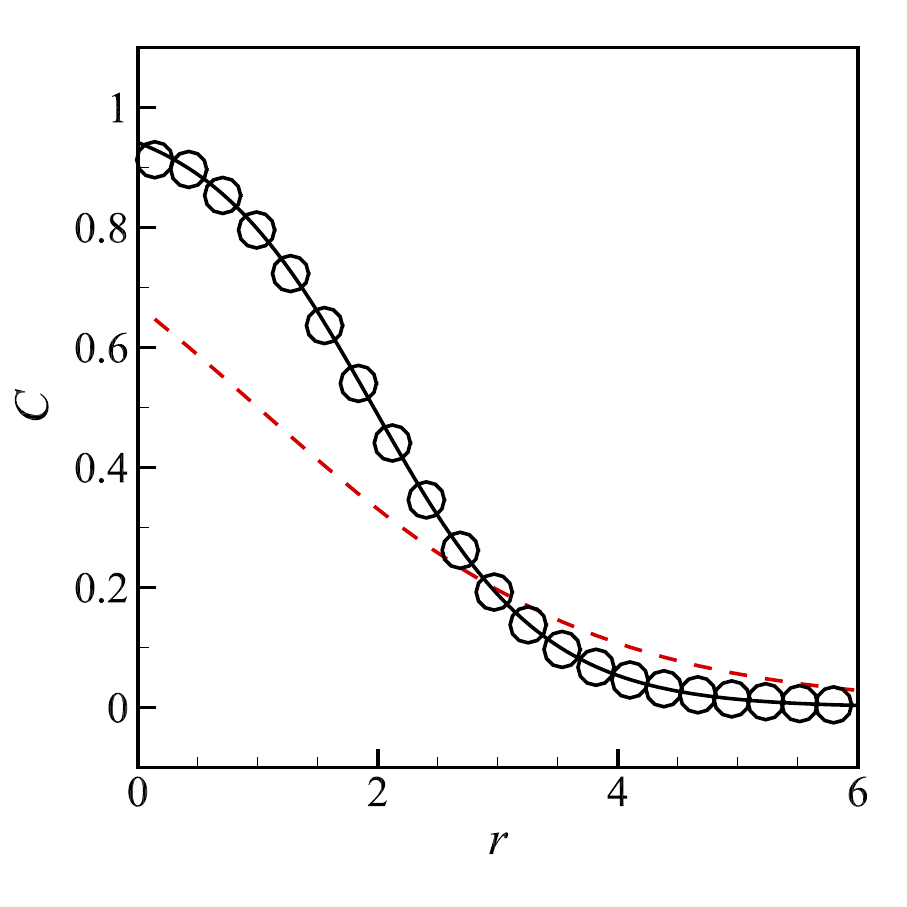}
\put(-180,160){($a$)}
\hspace{1cm}
\includegraphics[width=6cm]{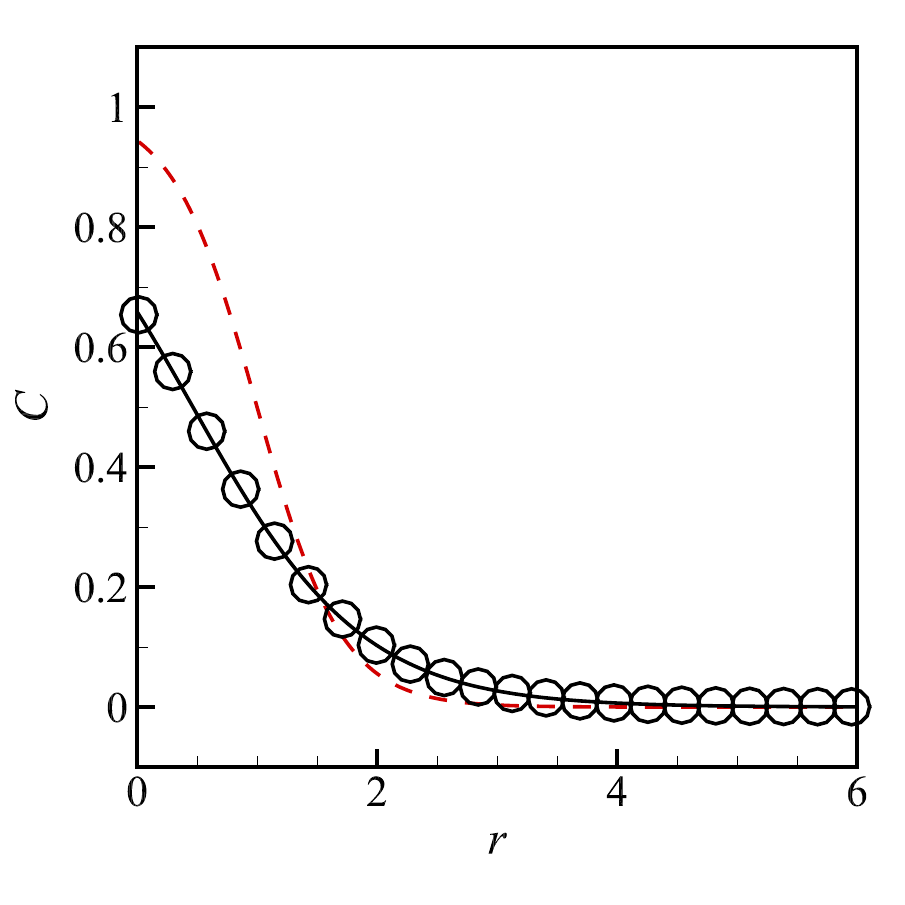}
\put(-180,160){($b$)}
\hspace{1cm}
\caption{Comparison of interface profile between the numerical results (symbols) and the equilibrium state (solid line) for a droplet with $Cn=0.5$ at $t=2$ with different initial interface profile: ($a$) stretched interface and ($b$) compressed interface. Note that the dashed lines represent the initial interface profiles.}
\label{case01.CompressStretch}
\end{figure}

We first check whether the interface can maintain the equilibrium state after long-term evolution, given an initial equilibrium profile for an under-resolved interface. Fig.~\ref{case01.Unsolve}($a$) shows the results of two typical cases: $Cn=0.1$ and $Cn=1$; in these two cases, the interface thickness is close to $R$ at $Cn=0.1$, and much larger than $R$ at $Cn=1$. We can see that regardless of the value of $Cn$, the numerical results of the interface profile agree well with the equilibrium state. Fig.~\ref{case01.Unsolve}($b$) shows the contours of the $C$ field at $Cn=1$ in the domain, and they virtually overlap with the theoretical prediction of the equilibrium profile. Then, we check whether the interface can recover the equilibrium state if under-resolved interfaces are initially out of equilibrium. Fig.~\ref{case01.CompressStretch} shows the results at $Cn=0.5$, including a stretched interface (Fig.~\ref{case01.CompressStretch}$a$) and a compressed interface (Fig.~\ref{case01.CompressStretch}$b$). The interface out of equilibrium is initialized by the formula $C=0.5+0.5\tanh(\mathcal{R} z/(2\sqrt{2}\epsilon))$, and specifically, $\mathcal{R}=0.5$ in Fig.~\ref{case01.CompressStretch}($a$) and $\mathcal{R}=2.0$ in Fig.~\ref{case01.CompressStretch}($b$). From the comparison between the numerical results and the equilibrium state, it is clear that the equilibrium interface profile is accurately restored in both cases by the improved PF model.

\begin{figure}[!t]
\centering
\includegraphics[width=14cm]{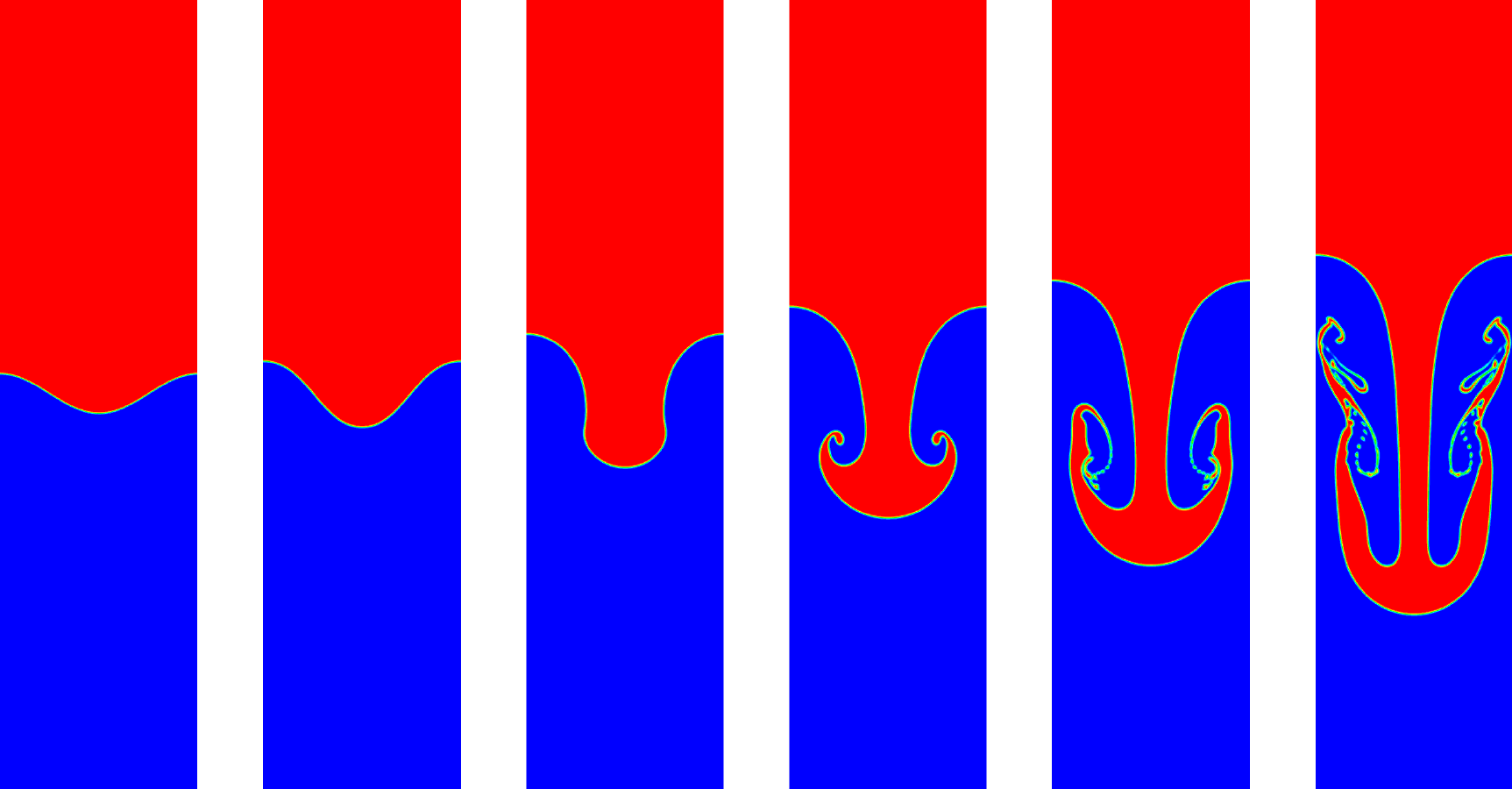}
\caption{Rayleigh–Taylor instability simulation with $A_t=0.5$ and $Re=3000$ at different times: $t=0$, $0.5$, $1.0$, $1.5$, $2.0$ and $2.5$ from left to right.}
\label{case1.RT_FlowProcess}
\end{figure}
\begin{figure}[!t]
\centering
\includegraphics[width=7cm]{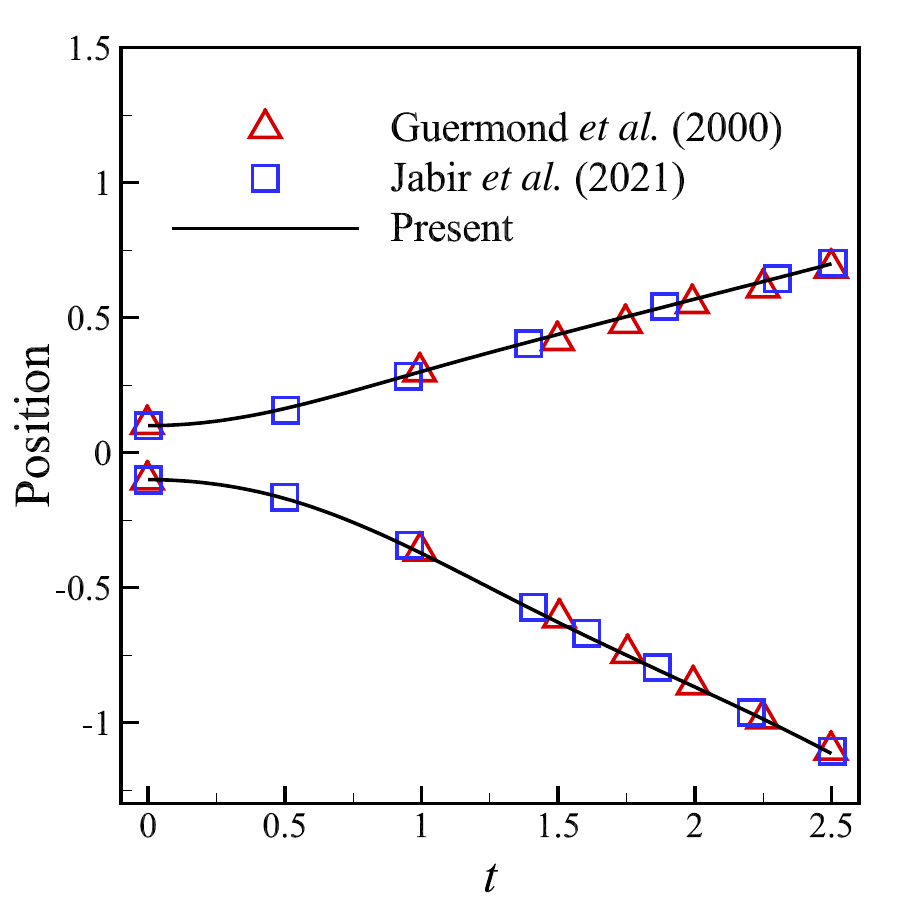}
\caption{The $y$-coordinate of tip of the falling and rising fluid versus time: solid lines denote the present solution, the triangles represent the solution of  Guermond and Quartapell~\cite{Guermond2000jcp} and the filled triangles that of Al-Salami et al~\cite{Jabir2021jcp}.}
\label{case1.RT_Iterface_UpDown}
\end{figure}

\subsection{Rayleigh–Taylor instability}
\label{RTinstability}
We consider here a two-dimensional Rayleigh-Taylor problem, in which a heavy fluid $1$ is placed on a light fluid $2$. The difference in density can be represented by the Atwood number $A_t=(\rho_1-\rho_2)/(\rho_1+\rho_2)$, where $\rho_i$ represents the density of fluid $i$ (=1 or 2). The fluids are assumed to have the same viscosity $\mu$, and the effect of surface tension at the interface between them is neglected. The computational domain is a rectangular one: $0\le x \le L$ and $0\le y\le 4L$, and the boundary conditions are set as follows: solid walls at the top and bottom, and periodic conditions at the left and right. Initially, the interface position is prescribed as $y(x)=2L+0.1L\cos(2\pi x/L)$, corresponding to a planar interface superimposed with a cosine perturbation. The initial numerical setup can be seen in the first panel in Fig.~\ref{case1.RT_FlowProcess}. The evolution of the perturbation on the fluid-fluid interface is primarily driven by gravitational acceleration. Therefore, the characteristic velocity $U$ is chosen as $\sqrt{gL}$, and consequently, the Reynolds number is defined as $Re=\rho_1 L^{3/2} g^{1/2}/\mu$ and Froude number $Fr=U^2/(gL)$ has a constant value $Fr=1$. The absence of surface tension allows for a thin interface in the simulation, e.g. by setting the Cahn number to $Cn=0.3 h$, where $h=0.0050$ is used in this study. To verify the accuracy of the present method, we adopt the same parameters as those used in previous studies~\cite{Guermond2000jcp, Jabir2021jcp}: $A_t=0.5$ and $Re=3000$. 

Figure~\ref{case1.RT_FlowProcess} illustrates interface snapshots in the simulation with $Pe=0.01Cn^{-2}$ at different times, which are made dimensionless by $\sqrt{L/(gA_t)}$. With the growth of the interface perturbation, the interface rolls up, forming two counter-rotating vortices. The evolution of interface structures qualitatively agrees well with previous research~\cite{Tryggvason1988jcp, Sheu2009jcp, Huang2020jcp}. For a quantitative comparison, we track the temporal evolution of the fronts of the descending heavy fluid and the ascending light fluid respectively, and present the results in Fig.~\ref{case1.RT_Iterface_UpDown}, along with the benchmark data~\cite{Guermond2000jcp, Jabir2021jcp}. Clearly, an excellent agreement has been achieved.

\begin{figure}[!t]
\centering
\includegraphics[width=14cm]{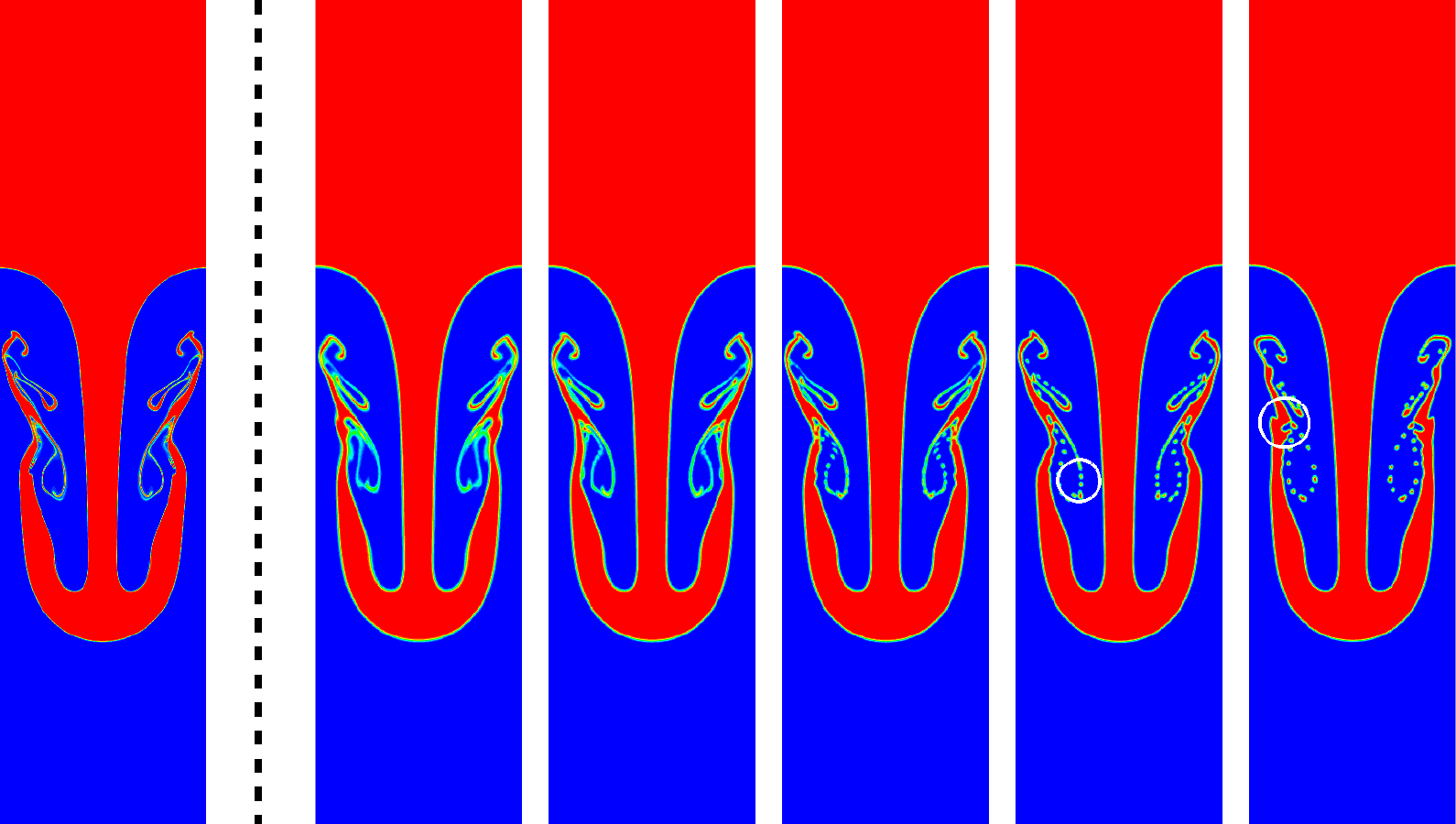}
\put(-398,230){($a$)} \put(-314,230){($b$)} \put(-250,230){($c$)} \put(-186,230){($d$)}
\put(-122,230){($e$)} \put(-58,230){($f$)}
\caption{Comparison of the interface shapes for different $Pe$ number at $t=2.5$. The panel ($a$) shows the result using the CH-PF method~\cite{Ding2007jcp} with $h=0.00125$, which serves as a reference solution. The panels from ($b$) to ($f$) are the results using the improved PF method obtained with $h=0.005$ and $Pe=0.1Cn^{-2}$, $0.03Cn^{-2}$, $0.01Cn^{-2}$, $0.003Cn^{-2}$ and $0.001Cn^{-2}$, respectively.}
\label{case1.RT_Different_Pe}
\end{figure}
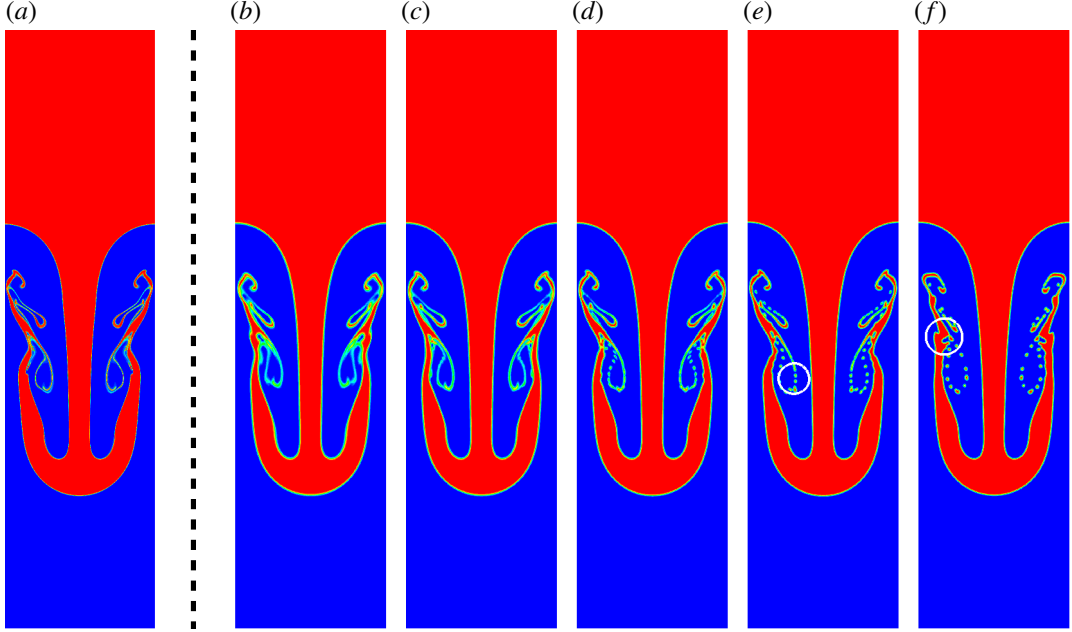

To determine the appropriate value of $\lambda$ in the P\'eclet number, we investigate the effect of the varying $Pe$ on the interface evolution. The numerical results at $t=2.5$ using the CH-PF model~\cite{Ding2007jcp} on a fine mesh ($h=0.00125$) is adopted as the benchmark solution, which shows the sophisticated interface structures Fig.~\ref{case1.RT_Different_Pe}(a). Accordingly, we simulate the same case on a relatively coarse mesh ($h=0.005$) using the improved PF model for a wide range of $Pe$: $0.01Cn^{-2}$, $0.03Cn^{-2}$, $0.01Cn^{-2}$, $0.003Cn^{-2}$ and $0.001Cn^{-2}$, respectively. The snapshots of the Rayleigh-Taylor instability at $t=2.5$ are shown in Figs.~\ref{case1.RT_Different_Pe}(b)-\ref{case1.RT_Different_Pe}(f). We can see that large interface structures such as the growth of the interface fronts are more or less the same, suggesting that they are not sensitive to the value of $Pe$. However, small interface structures such as the rolling up of the interfaces are smeared at relatively large $Pe$, e.g. $Pe=0.1Cn^{-2}$ in Fig.~\ref{case1.RT_Different_Pe}(b), and becomes sharper with the decrease of $Pe$, e.g. $Pe=0.1Cn^{-2}$ in Figs.~\ref{case1.RT_Different_Pe}(d) and \ref{case1.RT_Different_Pe}(e). On the other hand, too small $Pe$ might give rise to unphysical changes of interface structures, e.g. by comparing $Pe=0.001Cn^{-2}$ in Fig.~\ref{case1.RT_Different_Pe}(f) with the benchmark solution. This unphysical changes of interfaces are attributed to the excessive correction of the diffusion term in the improved PF model. From the numerical results of the Rayleigh-Taylor instability problem with various $Pe$, we conclude that the appropriate value of $L/(\lambda U)$ ranges from $0.003$ to $0.01$, so as to maintain interface profile and avoid unphysical changes of interfaces at the same time.

\subsection{Rising bubbles in water}
\label{risingbubble}

\begin{figure}[!t]
\centering
\includegraphics[width=7cm]{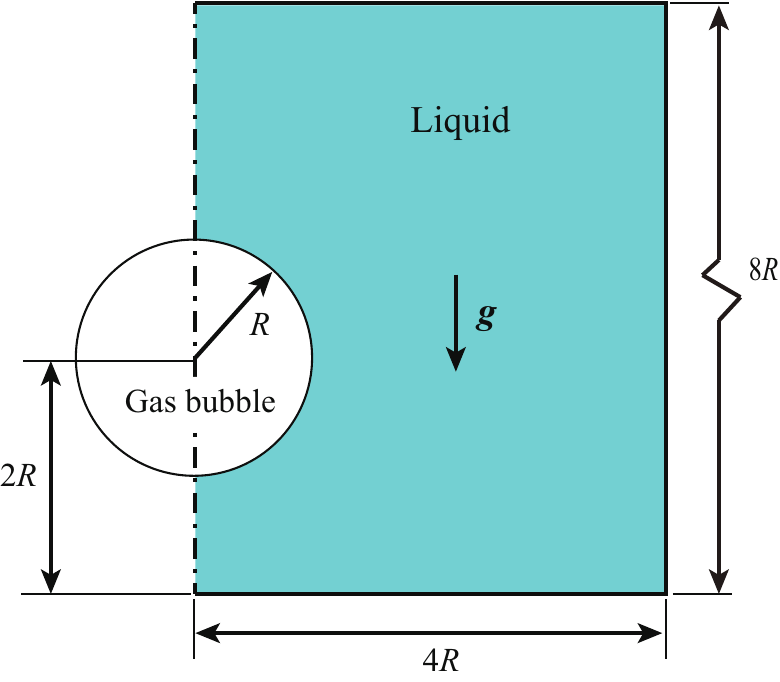}
\caption{Configuration for the rising bubble with buoyancy.}
\label{case2.RisBub_Sketch}
\end{figure}

We simulate axisymmetric rising of bubbles in water, to evaluate the performance of the method for two-phase flows with large density ratio ($\rho_2/\rho_1=0.001$) and viscosity ratio ($\mu_2/\mu_1=0.01$), where the subscripts $1$ and $2$ denote liquid and gas, respectively. The large density ratio between liquid and gas causes significant difference in buoyancy force across the diffuse interface of the bubble, thereby resulting in noticeable difference in flow acceleration in the interface. In particular, the strain flow at the bottom of the bubble tends to stretch the interface, thereby progressively increasing the local interface thickness, which makes maintaining the interface at equilibrium a tough job. The numerical setup of axisymmetric simulations is shown in Fig.~\ref{case2.RisBub_Sketch}: The computational domain is $(0, 4R)\times(0, 8R)$, with a bubble of radius $R$ initially located at $(0, 2R)$. Symmetric condition is imposed at the left boundary, while slip boundary condition is enforced at the other boundaries. We choose the same parameters as in Ding et al.~\cite{Ding2007jcp}, i.e., $Re$($=\rho_1 g^{1/2}R^{3/2}/\mu_1$)=100 and $Bo$($=\rho_1 g R^{2}/\sigma$)=200.

\begin{figure}[!t]
\centering
\includegraphics[width=5cm]{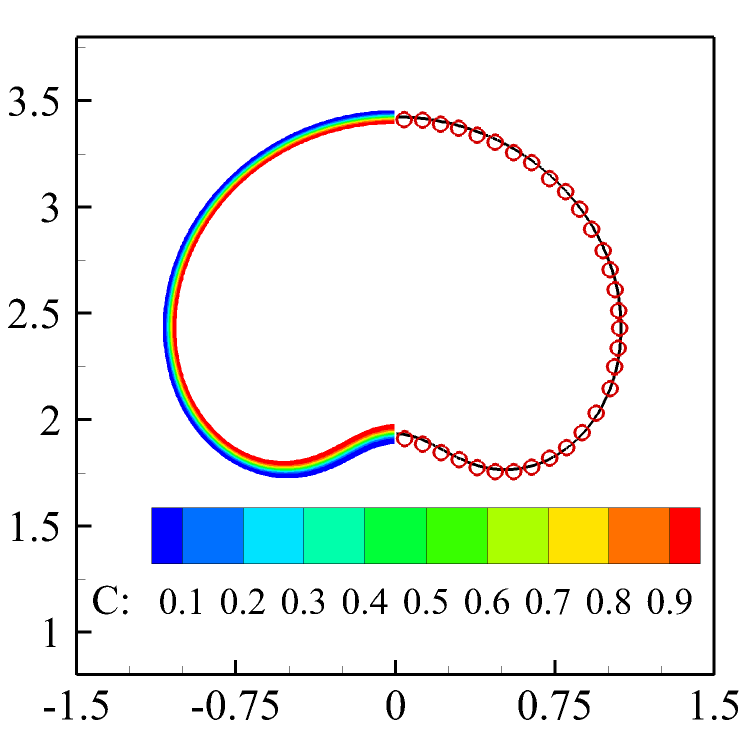}
\hspace{.5cm}
\includegraphics[width=5cm]{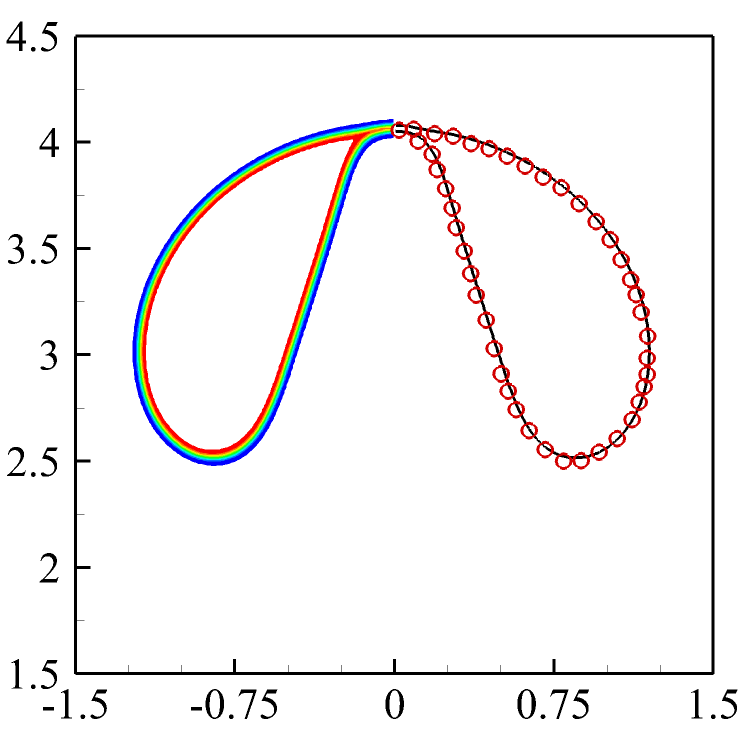}
\hspace{.5cm}
\includegraphics[width=5cm]{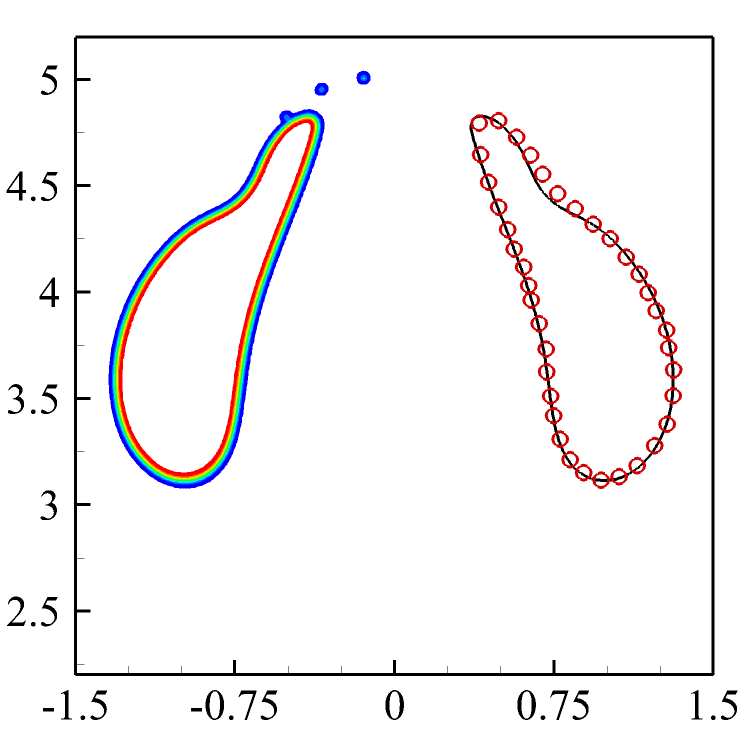}
\put(-440,120){($a$)}\put(-280,120){($b$)} \put(-120,120){($c$)}
\caption{Snapshots of the rising bubble at times $t=0.8$ (a), 1.6 (b) and 2.4 (c) with $h = 1/100$, $Cn = 0.5h$ and $Pe = 0.01Cn^{-2}$. The left half of each figure shows the contours of $C$ from 0.05 to 0.95, and the right half shows the comparison with the previous study~\cite{Ding2007jcp} (red circles).}
\label{case2.RisBub_FlowProcess}
\end{figure}

Figure~\ref{case2.RisBub_FlowProcess} illustrates snapshots of the rising bubble with $h=0.01$, $Cn=0.5h$ and $Pe=0.01Cn^{-2}$, in terms of the contours of $C$. Specifically, the snapshots at times $t=0.8$, $1.6$ and $2.4$ correspond to three different stages of bubble dynamics: deformation, onset of breakup and breakup. All the results are in good agreement with previous research~\cite{Ding2007jcp} with respect to the interface shape (represented by the contour $C=0.5$). Moreover, the bubble breakup occurs at $t = 1.61$ and $z = 4.10$, which is consistent with the observations in~\cite{Ding2007jcp} ($t = 1.61$ and $z = 4.10$) and~\cite{Sussman1997jfm} ($t = 1.60$ and $z = 4.05R$). We also note that there is no distortion of interface profile and the interface thickness remains nearly uniform along the bubble's surface. The under-resolved bubbles arising from the breakup in Fig.~\ref{case2.RisBub_FlowProcess}($c$) do not dissolve in the surrounding liquid, which is distinct from simulations using the CH-PF model~\cite{Yue2007jcp}.

\begin{figure}[!t]
\centering
\includegraphics[width=10cm]{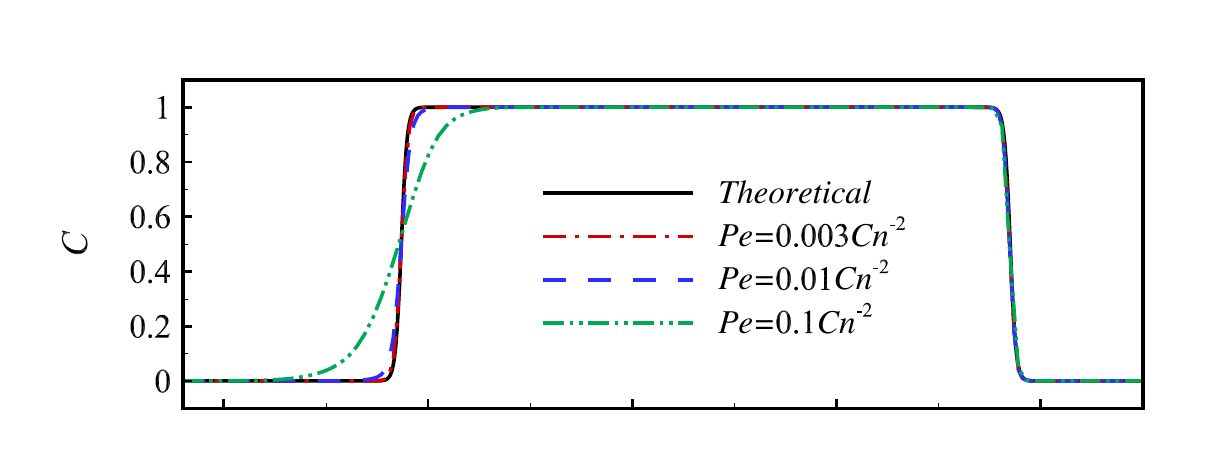}
\put(-285,85){($a$)}\\
\includegraphics[width=10cm]{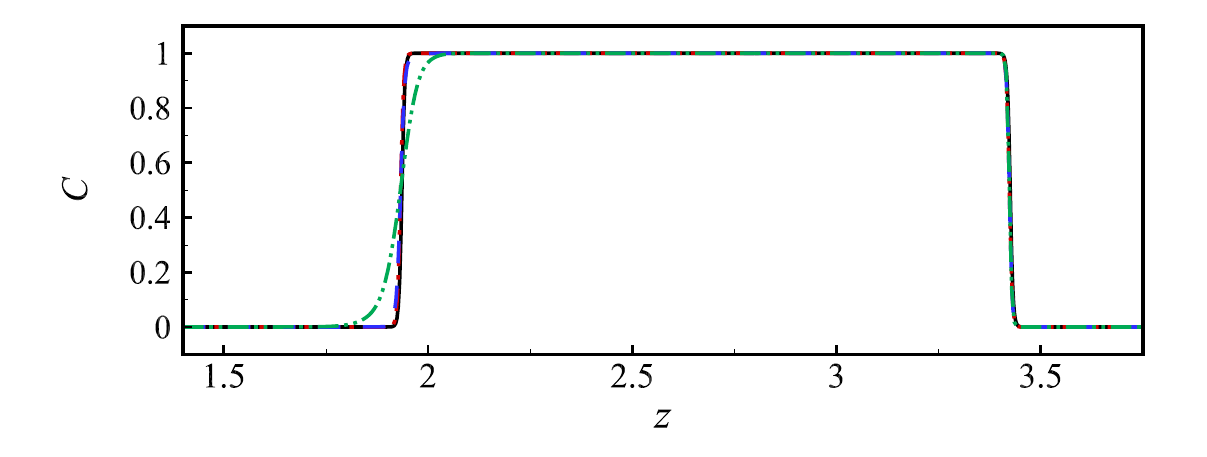}
\put(-285,95){($b$)}
\caption{Distribution of $C$ along the symmetry axis at $t=0.8$ with various $Pe$, $h$ and $Cn$: (a) $h=1/100$ and $Cn = 0.50h$ and (b) $h=1/300$ and $Cn = 0.75h$.}
\label{case2.RisBub_Different_Pe}
\end{figure}

To assess the sensitivity of the present model to the value of $Pe$, simulations are conducted for the same case, but with various $Pe$: $0.003Cn^{-2}$, $0.01Cn^{-2}$ and $0.1Cn^{-2}$, respectively. Figure~\ref{case2.RisBub_Different_Pe} illustrates the distribution of $C$ along the symmetrical axis at $t=0.8$, i.e. the interface profile in Fig.~\ref{case2.RisBub_FlowProcess}($a$); more specifically, $h=1/100$ and $Cn = 0.50h$ in Fig.~\ref{case2.RisBub_Different_Pe}($a$) and $h=1/300$ and $Cn = 0.75h$ in Fig.~\ref{case2.RisBub_Different_Pe}($b$). From the comparison between numerical results and the theoretical prediction of interface profile in equilibrium (where the interface position is evaluated by $C=0.5$), it is evident that the P\'eclet number has a notable influence on the interface profile, although the interface positions of the results are more or less the same ($z\approx 1.93$ and $3.42$ respectively). At relatively large $Pe$ ($=0.1Cn^{-2}$), significant stretching of interface profile is observed near the bottom of the bubble (around $z=1.93$), indicating that the diffusion flux in the improved PF model is not sufficiently strong to restore the interface to the equilibrium state. With the decrease of $Pe$ ($=0.01Cn^{-2}$ and $0.003Cn^{-2}$), the interface profiles nearly coincide with the theoretical prediction, which is true for different $h$ (and $Cn$). It is noteworthy that the appropriate range of $Pe$ for simulating the rising bubbles is consistent with those obtained from the Rayleigh-Taylor instability problem, implying that the effective range of $Pe$ ($0.01Cn^{-2}\ge Pe \ge 0.003Cn^{-2}$) could be universal. 

\begin{figure}[!t]
\centering
\includegraphics[width=6cm]{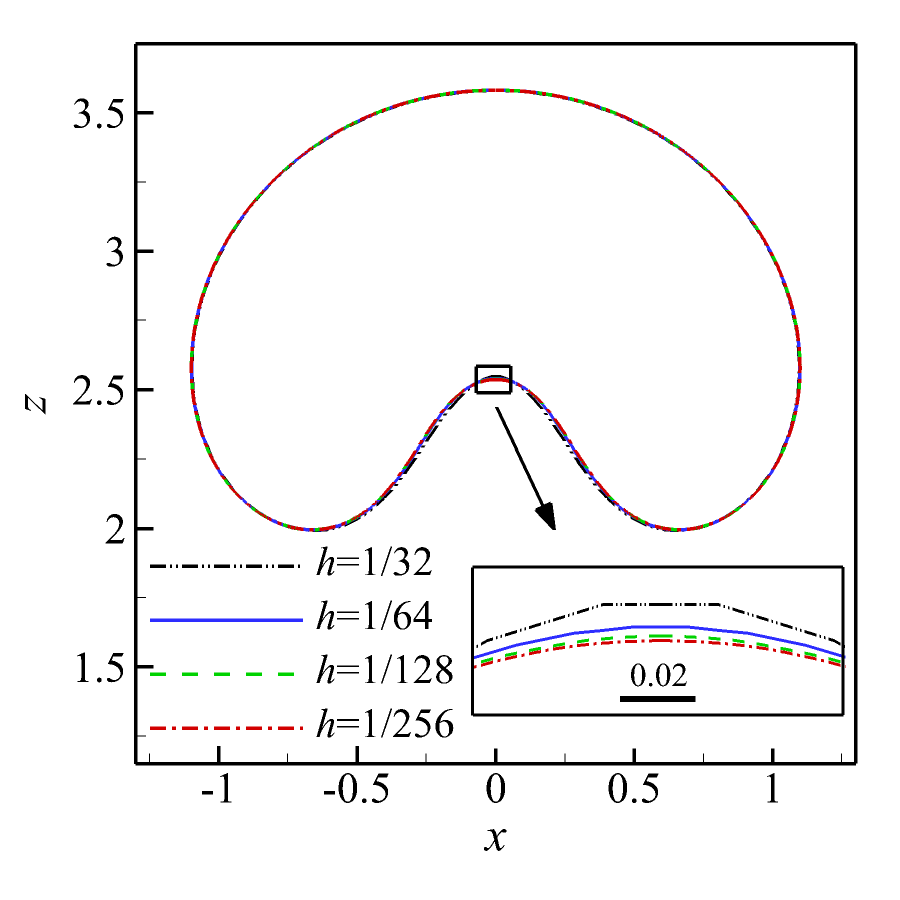}
\put(-180,160){($a$)}
\hspace{1cm}
\includegraphics[width=6cm]{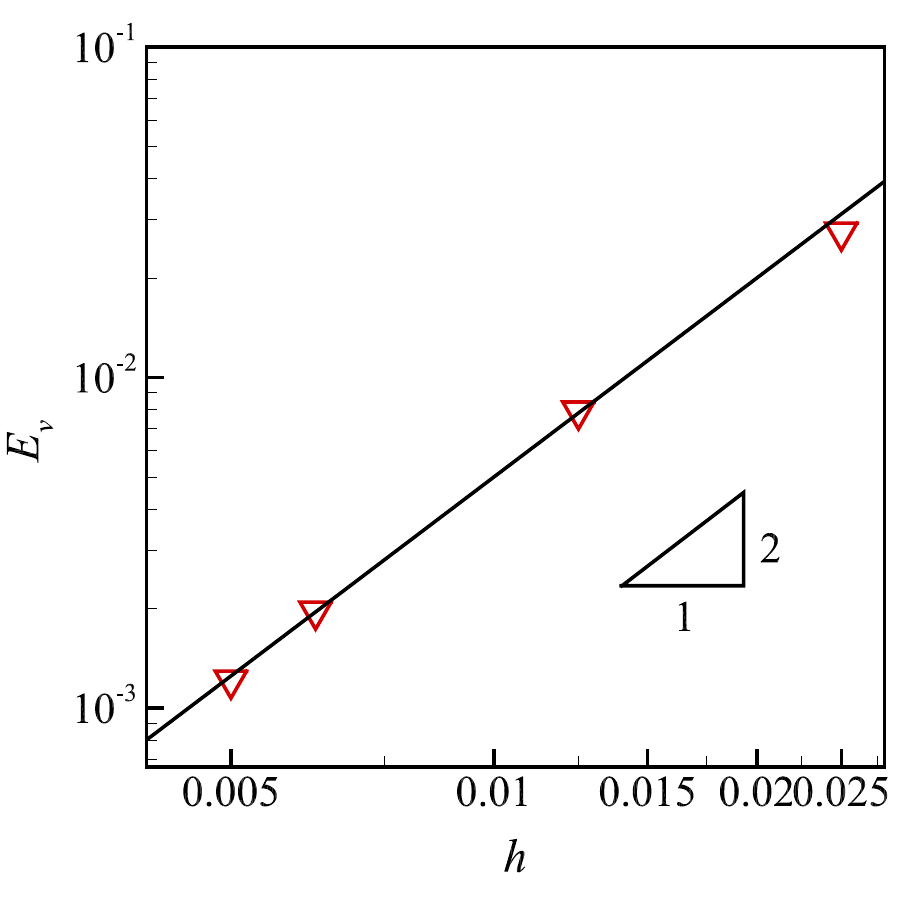}
\put(-180,160){($b$)}
\caption{(a) Convergence study in terms of interface profile at $t=1$. The inset shows a zoom-in view of the bubble bottom. (b) Convergence study in terms of $E_v$.}
\label{case2.RisBub_Convergence}
\end{figure}

Figure~\ref{case2.RisBub_Convergence} shows the convergence study with $Cn=0.5h$ and $Pe=0.01Cn^{-2}$ on various meshes: $h = 1/32$, $1/64$, $1/128$ and $1/256$, with respect to the bubble shape (Fig.~\ref{case2.RisBub_Convergence}$a$) and the vertical velocity at the bubble bottom (Fig.~\ref{case2.RisBub_Convergence}$b$) at $t=1$. The bubble shapes on the different meshes are apparently overlapped, and the zoom-in view of the bubble bottom at the symmetric axis shows that the interface converges with mesh refinement. A quantitative study of the convergence rate with mesh refinement is shown in Fig.~\ref{case2.RisBub_Convergence}($b$), in terms of the vertical velocity at the bubble bottom. The results on a fine mesh ($h = 1/320$) are chosen as a reference to calculate the numerical error $E_v$. We can see that a convergence rate of $2$ has been achieved. We also check the total mass of the bubble by integrating the $C$ field over the whole domain, and find that the mass loss is of order of machine accuracy ($10^{-16}$). This can be expected because of the finite volume discretization of the improved PF model. 

It should be emphasized that the same formula of dimensionless mobility $1/Pe=100Cn^2$ is implemented in the simulations in Fig.~\ref{case2.RisBub_Convergence}. Consequently, the value of $1/Pe$ decreases from $2.4\times10^{-2}$ to $3.8\times10^{-4}$ with mesh refinement from $1/32$ to $1/256$, spanning two orders in magnitude. This is very different from previous studies~\cite{Chiu2011jcp, Mirjalili2020jcp, Mirjalili2024jcp}, in which the mobility is recommended to take the value of dimensionless maximum velocity $\sim O(1)$. 

\subsection{Droplet migration due to the Marangoni effect}
\label{Marangoni}

\begin{figure}[!t]
\centering
\includegraphics[width=6cm]{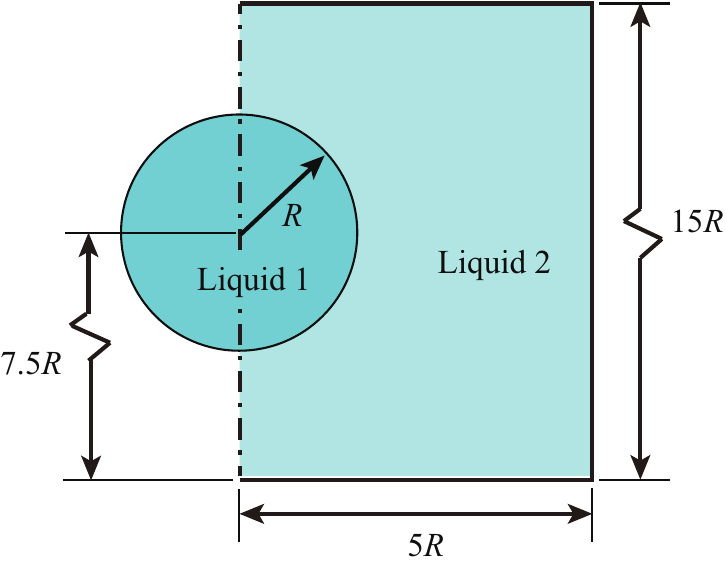}
\caption{Configuration for droplet migration due to the Marangoni effect.}
\label{case3.Marangoni_Sketch}
\end{figure}

Marangoni effect arising from an uneven distribution of surface tension coefficient can induce interfacial flows and cause droplet migration. To simulate such flows, it is crucial to accurately calculate the normal and tangential components of the surface tension, which poses a challenge for interface capturing methods. We consider here the migration of a viscous droplet with radius $R$ surrounded by ambient liquid with a constant linear temperature gradient $T_{z}$, which has been widely adopted for verification of surface tension model~\citep{Muradoglu2008jcp, Teigen2011jcp, Stricker2017jcp, Abu2018jcp, Seric2018jcp}. Under the assumption of Stokes flow, the terminal migration velocity of droplet can be obtained as~\citep{Young1959jfm}
\begin{equation}
    V_{YGB}=\frac{-2\sigma_T T_{z} R}{6 \mu_1 +9 \mu_2 },
\label{V_{YGB}}
\end{equation}
where $\sigma_T$ is the gradient of the surface tension coefficient with respect to the temperature. $\mu_1$ and $\mu_2$ are the viscosity of droplets and ambient fluids, respectively. The numerical setup is shown in Fig.~\ref{case3.Marangoni_Sketch}, where a viscous droplet of radius $R$ initially rests inside another static fluid, with its center located at ($0, 7.5R$). The simulations are axisymmetric and in a domain of $5R\times15R$, with symmetric condition imposed at the left boundary and slip conditions at the other boundaries. The droplet and the ambient fluid are assumed to have the same density ($\rho=\rho_1=\rho_2$) and viscosity ($\mu=\mu_1=\mu_2$). A linear temperature field is imposed in the $z$-direction. The dimensionless parameters of this problem are defined as: Reynolds number $Re=\rho V_{YGB}R/\mu$ and capillarity number $Ca=\mu V_{YGB}/\sigma_0$, where $\sigma_0$ is the surface tension coefficient of the droplet at the reference temperature. Here, we use the same parameters as Teigen
et al.~\cite{Teigen2011jcp}, i.e. $Re=0.0889$ and $Ca=0.0089$. In the numerical simulation, we choose $h = 1/50$, $Cn=0.5h$, and $Pe=0.007Cn^{-2}$.

\begin{figure}[!t]
\centering
\includegraphics[width=6cm]{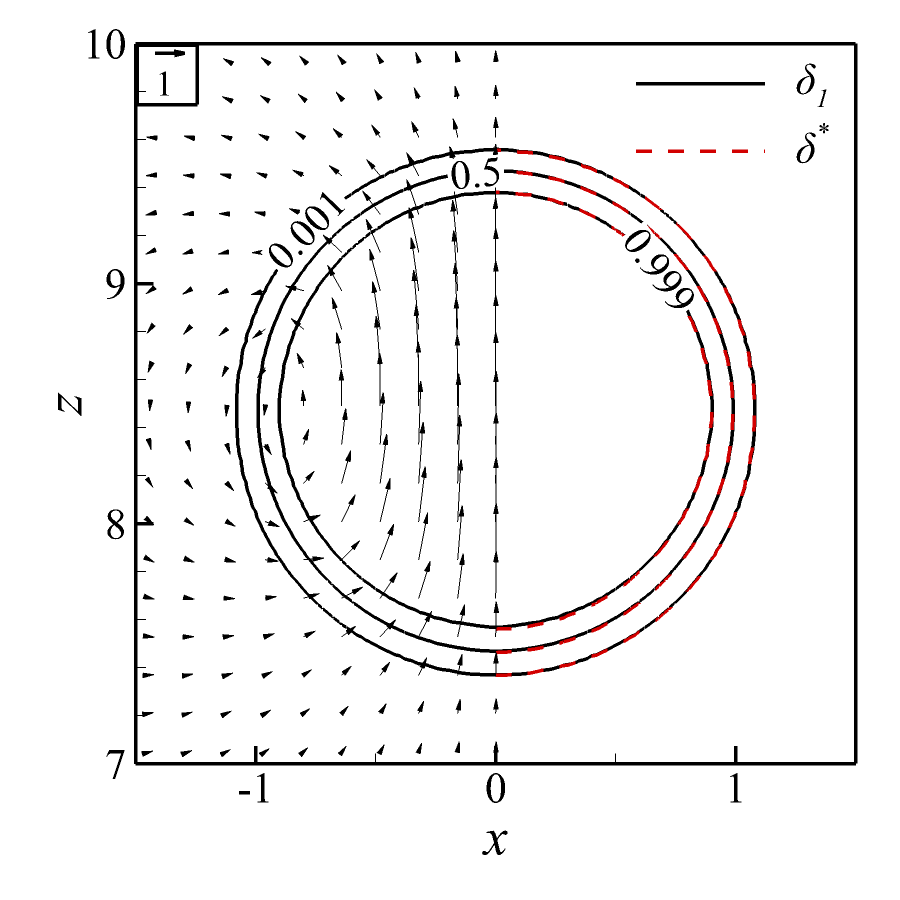}
\put(-175,155){($a$)}
\hspace{1cm}
\includegraphics[width=6cm]{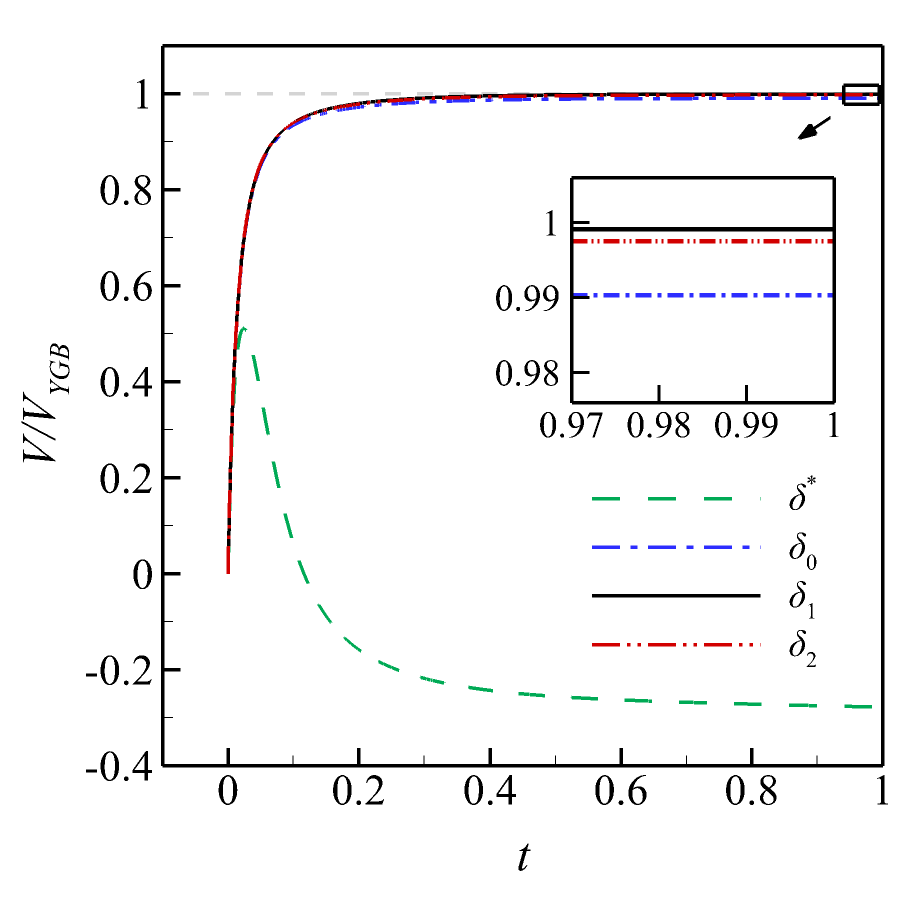}
\put(-175,155){($b$)}
\caption{(a) Droplet migration due to the Marangoni effect at $t=1$ in terms of velocity vectors (arrows) and interface contours of $C=0.001$, $0.5$ and $0.999$. The solid and dashed lines represent results with $\delta_1$ and $\delta^{*}$ for the surface tension calculation, respectively. The contours of $\delta^{*}$ are shifted and aligned with $\delta_1$ for the convenience of comparing the interface contours. (b) Temporal evolution of droplet migration velocity using different $\delta$ functions in the surface tension calculation.}
\label{case3.Marangoni}
\end{figure}

Figure~\ref{case3.Marangoni}(a) shows the flow field and contours of $C$ at time $t=1$. It can be observed that an upward flow field is formed inside the droplet due to the surface tension gradient at the interface, and the whole flow pattern is qualitatively the same as that of Young et al.~\cite{Young1959jfm}. From the contour of the $C$, it can also be seen that the interface remains uniform during the movement process, and the interface will not be in a local non-equilibrium state, which is important for the accurate calculation of the normal direction and curvature of the interface. Quantitatively, the drop migration velocity $V$ is obtained by
\begin{equation}
V=\frac{\int_{0}^{15R}\int_{0}^{5R}C v(r, z) r drdz}{\int_{0}^{15R}\int_{0}^{5R}C r drdz},
\label{Velocity}
\end{equation}
where $v(r, z)$ represents the velocity in the vertical direction at coordinates $(r, z)$. The temporal evolution of the migration velocity of the drop is shown in Fig.~\ref{case3.Marangoni}(b). By employing various surface tension models, we observe that the final migration velocities of the drops consistently converge towards the theoretical solution, thereby validating the accuracy of these models. On the other hand, there are slight variations in how closely they approach the theoretical value. Specifically, the result obtained using $\delta_1$ exhibits the closest proximity to the theoretical solution, followed by the results obtained using $\delta_0$, while the one using $\delta_0$ yields the least accurate outcome. It demonstrates the effect of different surface tension thicknesses on the numerical results. That is, a thicker surface tension layer, as simulated with $\delta_0$, leads to a reduction in effective surface tension at the phase interface and consequently an underestimated droplet's migration velocity relative to the theoretical prediction. Conversely, when the surface tension layer is excessively thin (e.g. $\delta_2$), it effectively causes a decrease in the number of meshes resolving the surface tension. As a result, the velocity of droplet migration does not match the theoretical solution as closely as those obtained with $\delta_1$.

We compare the effect of different $\delta$ functions: $\delta_1$ in Eq.~(\ref{delta}) and $\delta^*=6\sqrt{2}\epsilon |\nabla C|^2$ on droplet migration. Fig.~\ref{case3.Marangoni} shows the results with respect to interface contours and the evolution of droplet migration velocity. We can see from Fig.~\ref{case3.Marangoni}($a$) that the interface profiles obtained by $\delta_1$ and $\delta^*$ are very similar, with the maximum difference of roughly $2$ mesh sizes regarding the interface shape ($C=0.5$). However, the small difference in interface profile can make a big difference in the migration velocity of the droplet owing to the use of different $\delta$ functions, as shown in Fig.~\ref{case3.Marangoni}($b$). For these $\delta$ functions that satisfy the fundamental property of the Dirac delta function, the migration velocity of the droplet eventually reaches the theoretical prediction (less than 1\% in relative error), and the result with $\delta_1$ is the closest one. By contrast, although the droplet using $\delta^*$ has the same evolution of velocity as those using  $\delta_n$ at the early stage of migration, it moves in an opposite direction when reaching the steady state.

\begin{figure}[!t]
\centering
\includegraphics[width=7cm]{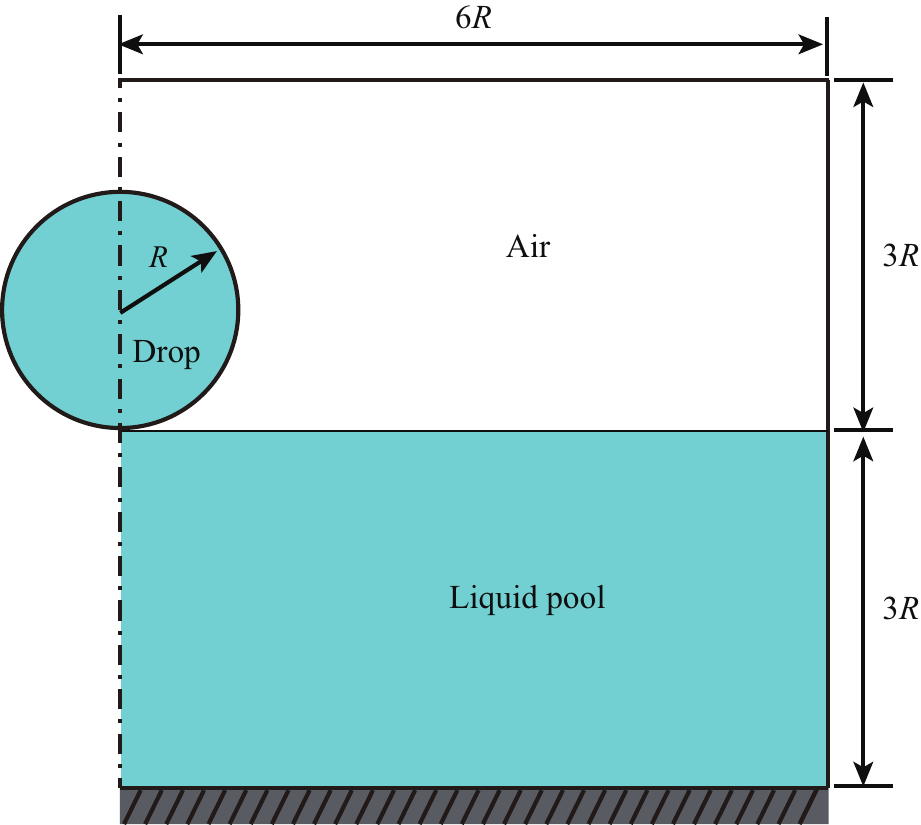}
\caption{Configuration for partial coalescence of a drop into a pool.}
\label{case4.DropPool_Sketch}
\end{figure}

\subsection{Partial coalescence of a drop with a pool}
\label{coalescence}

We simulate the dynamic process of partial coalescence of a drop with a pool, which has been extensively investigated by experiments~\citep{charles1960jcs,bach2004jfm, Blanchette2006np, Alhareth2020pof} and numerical simulations~\citep{blanchette2009jfm, Ding2014jcp, pang2024cf}. The setup of the simulation is shown in Fig.~\ref{case4.DropPool_Sketch}. A drop of radius $R$ is gently placed on the surface of a pool of the same liquid. The radius and depth of the pool are $6R$ and $3R$, respectively. The computational domain is axisymmetric and of size $6R\times 6R$. The following boundary conditions are applied: symmetric condition at the left boundary, solid-wall condition at the lower boundary, and slip condition at the right and upper boundaries. The dynamics of the partial coalescence is characterized by the Ohnesorge number $Oh=\mu/\sqrt{\rho R\sigma}$, where $\rho$ ($=759$g/m$^3$) and $\mu$ ($=0.49$ mPa s) are the density and viscosity of the liquid, respectively; $\sigma$ ($=15.9$ mN/m) is the surface tension coefficient between liquid and surrounding gas. The density and viscosity of the gas to the liquid are fixed at $0.0015$ and $0.037$, respectively. In the following simulations, we consider a small drop ($R=0.2765$ mm) as in the experiment~\citep{Alhareth2020pof}, which corresponds to the case with $Oh=0.00848$. Also, because $R$ is much smaller than the capillary length ($=\sqrt{\sigma/(\rho g)}=1.46$ mm), the effect of gravity is ignored. In the present study, we choose $h=0.01$, $Cn=0.5h$ and $Pe=0.007Cn^{-2}$.

\begin{figure}[!t]
\centering
\includegraphics[width=12cm]{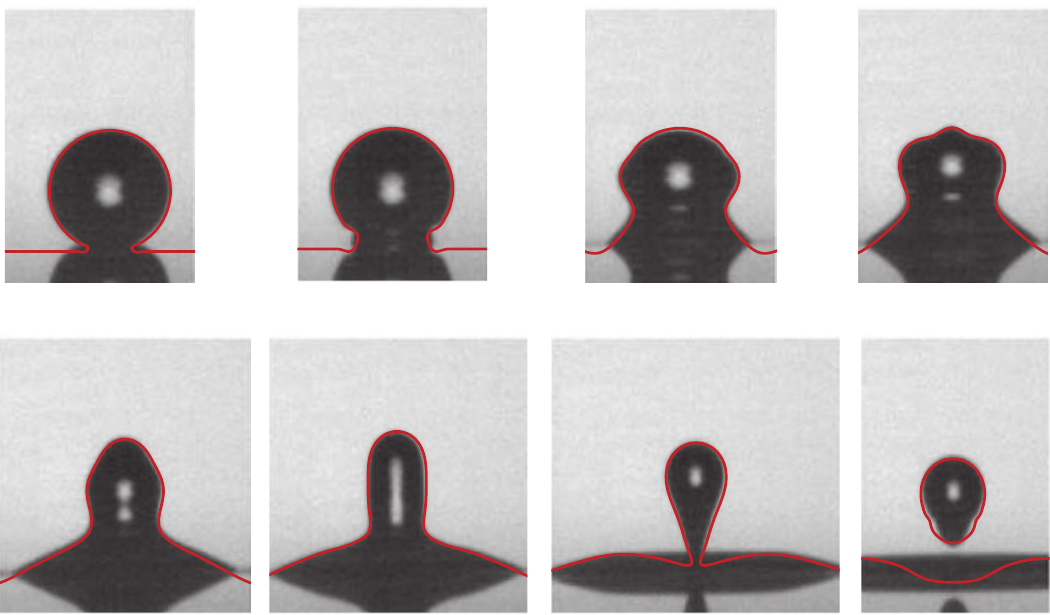}
\put(-340,200){$t=0$ms}\put(-245,200){0.08ms}
\put(-150,200){0.37ms}\put(-62,200){0.56ms}
\put(-340,92){0.90ms}\put(-255,92){1.10ms}
\put(-162,92){1.48ms}\put(-62,92){1.71ms}
\caption{The snapshots of interface shapes at different times in terms of the contour $C=0.5$, superimposed by the experimental images of Alhareth and Thoroddsen~\cite{Alhareth2020pof}. Note that $\delta_1$ is chosen in the surface tension calculation.}
\label{case4.DropPool_FlowProcess}
\end{figure}

\begin{figure}[!t]
\centering
\includegraphics[width=6cm]{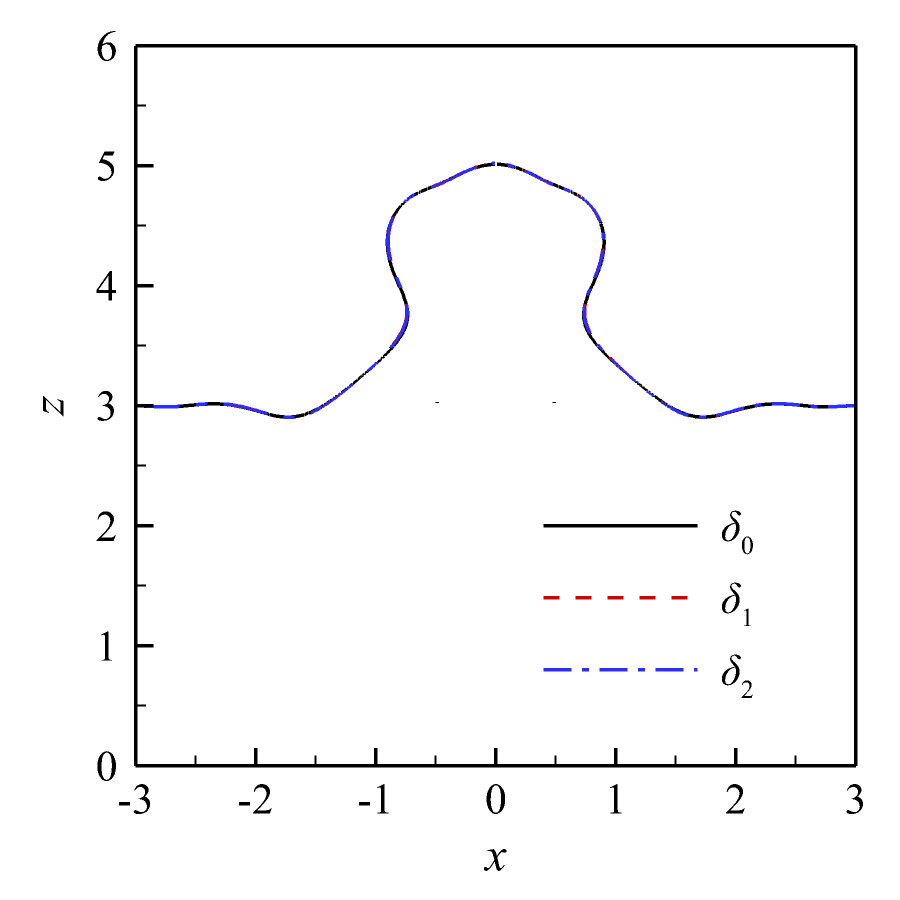}
\put(-180,160){($a$)}\hspace{1cm}
\includegraphics[width=6cm]{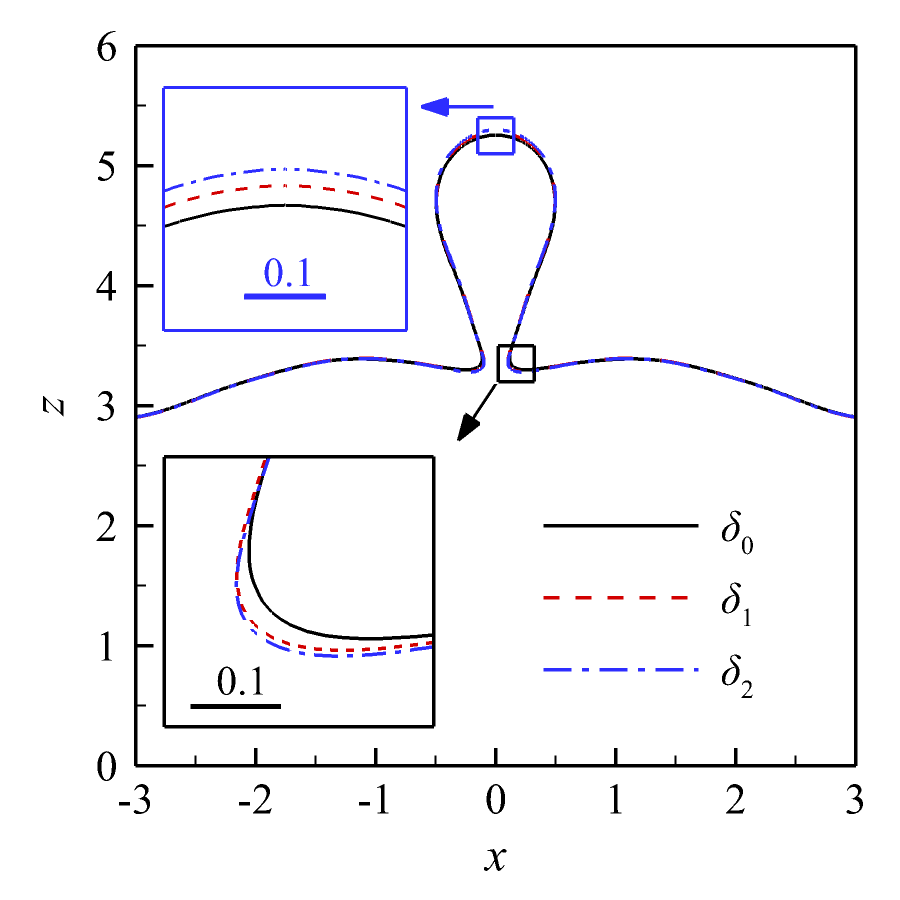}
\put(-180,160){($b$)}
\caption{Comparison of interfaces with different surface tension forms at two special moments: (a) $t=0.56$ ms when the capillary waves converge at the apex of the drop, and (b) $t=1.48$ ms when the pinch-off is imminent. The insets show a zoomed-in view of the interfaces.}
\label{case4.DropPool_T}
\end{figure}
Figure~\ref{case4.DropPool_FlowProcess} shows the snapshots of the interface shapes at different times, superimposed by the experimental images of Alhareth and Thoroddsen~\cite{Alhareth2020pof}. At the initial stage of partial coalescence, a neck is formed at the contact of the drop with the pool. This induces capillary waves travelling from the expanding neck towards the apex of the drop, resulting in significant deformation of the drop and shaping the drop into a column of liquid. At a late time, the liquid column ends up with a pinch-off, leaving a small satellite droplet above the interface. In general, the numerical results agree well with the experimental data. In particular, the pinch-off time ($t=1.48$ ms), pinch-off position and the size of the satellite drop are consistent with the experimental results.

Figure~\ref{case4.DropPool_T} shows the snapshots of the interfaces at $t=0.56$ ms and $1.48$ ms, for simulations with various delta functions ($\delta_0$, $\delta_1$ and $\delta_2$). The interfaces are generally distinguishable for all the delta functions, owing to the sufficiently fine resolution of the mesh size. However, Fig.~\ref{case4.DropPool_T}($b$) shows that the choice of the delta function could affect the results in the interface region with high curvature; specifically, compared with $\delta_1$ and $\delta_2$, $\delta_0$ tends to produce lower curvature, because of its less concentrated surface tension.

\subsection{Three-dimensional drop in simple shear flow}
\label{shear}

\begin{figure}[!t]
\centering
\includegraphics[width=4.5cm]{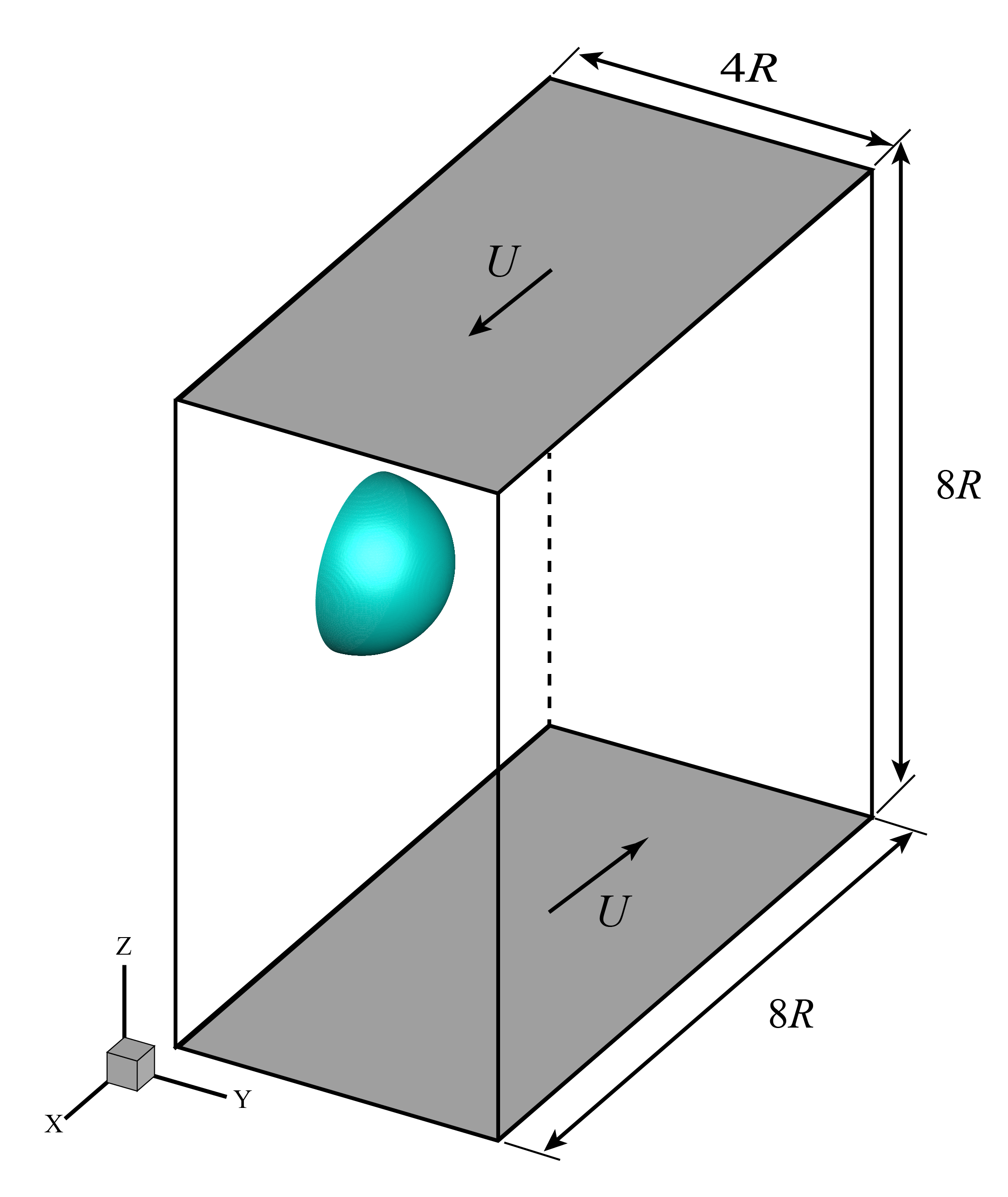}
\caption{Configuration for a three-dimensional drop in simple shear flow.}
\label{case5.ShearFlow_Sketch}
\end{figure}

We simulate the dynamics of a three-dimensional drop in shear flows with matched density and viscosity. The interplay among viscous force, surface tension and drop inertia dominates the drop dynamics and can be characterized by two dimensionless numbers: capillary number $Ca=\mu\dot{\gamma} R/\sigma$ and Reynolds number $Re=\rho\dot{\gamma}R^2/\mu$, where $R$ is the drop radius, $\dot{\gamma}$ the shear rate of the flow, $\rho$ and $\mu$ the density and viscosity of the drop (and the surrounding fluid), and $\sigma$ the surface tension coefficient. To quantify the drop deformation, a Taylor deformation parameter $D_T=(L-B)/(L+B)$ is employed, where $L$ and $B$ represent the lengths of the major and minor axes of the deformed ellipsoidal drop, respectively. Based on the assumptions of negligible inertia ($Re\ll 1$) and small drop deformation ($Ca\ll 1$), Taylor~\cite{Taylor1932} proposed a linear relationship between $D_T$ and $Ca$, and specifically, $D_T=(35/32)Ca$. This linear relation can be used to verify our three-dimensional simulations. 

Numerical setup is shown in Fig.~\ref{case5.ShearFlow_Sketch}. The drop with radius $R$ is initially positioned at $(4R, 0, 4R)$ in a computational domain $8R\times 4R\times 8R$. The top and bottom of the domain are two no-slip flat plates moving with a speed of $U=4\dot{\gamma}R$ in the opposite direction. Periodic boundary conditions are applied in the $y-z$ plane at $x=0$ and $x=8R$, while symmetric boundary conditions are used in the other directions. The height of the domain is chosen to be sufficiently large to minimize the effects of channel confinement~\cite{Janssen2010, Farokhirad2013ccp}. The simulations are performed on a uniform mesh of $320\times160\times320$, in which the initial radius of the drop is resolved by 40 mesh sizes, i.e. $h = 1/40$. The interface thickness is prescribed by setting $Cn=0.5h$, and the dimensionless mobility is set to $Pe=0.007Cn^{-2}$.

\begin{figure}[!t]
\centering
\includegraphics[width=6cm]{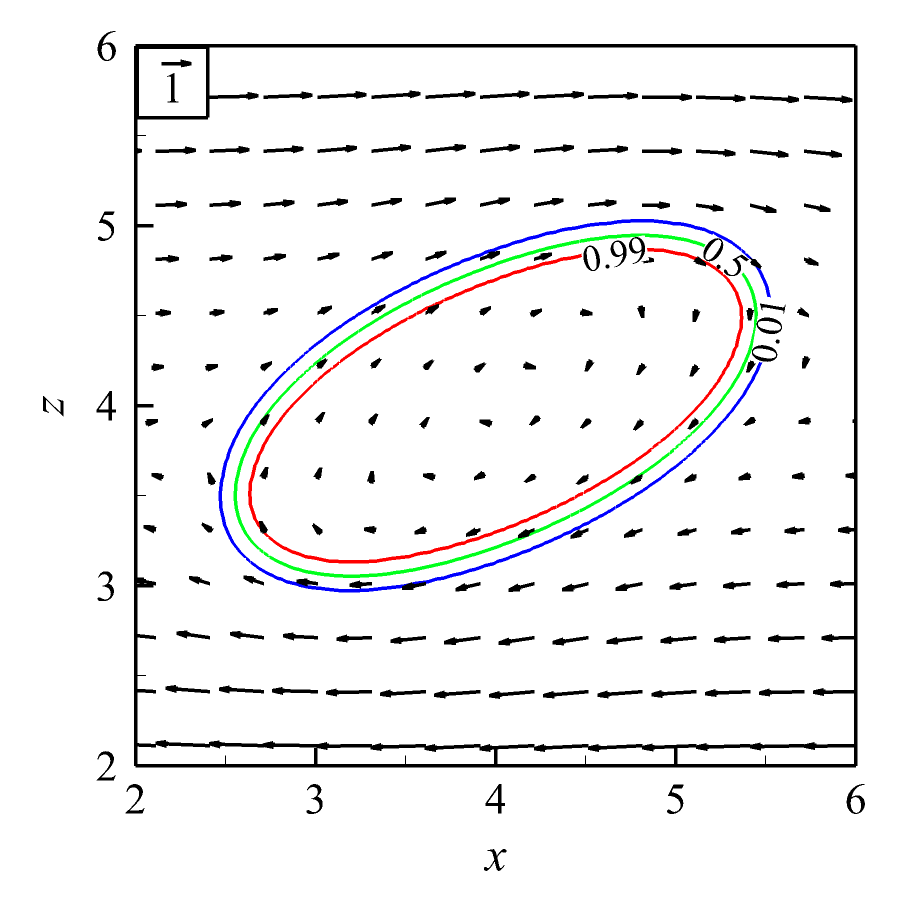}
\put(-180,158){($a$)}\hspace{1cm}
\includegraphics[width=6cm]{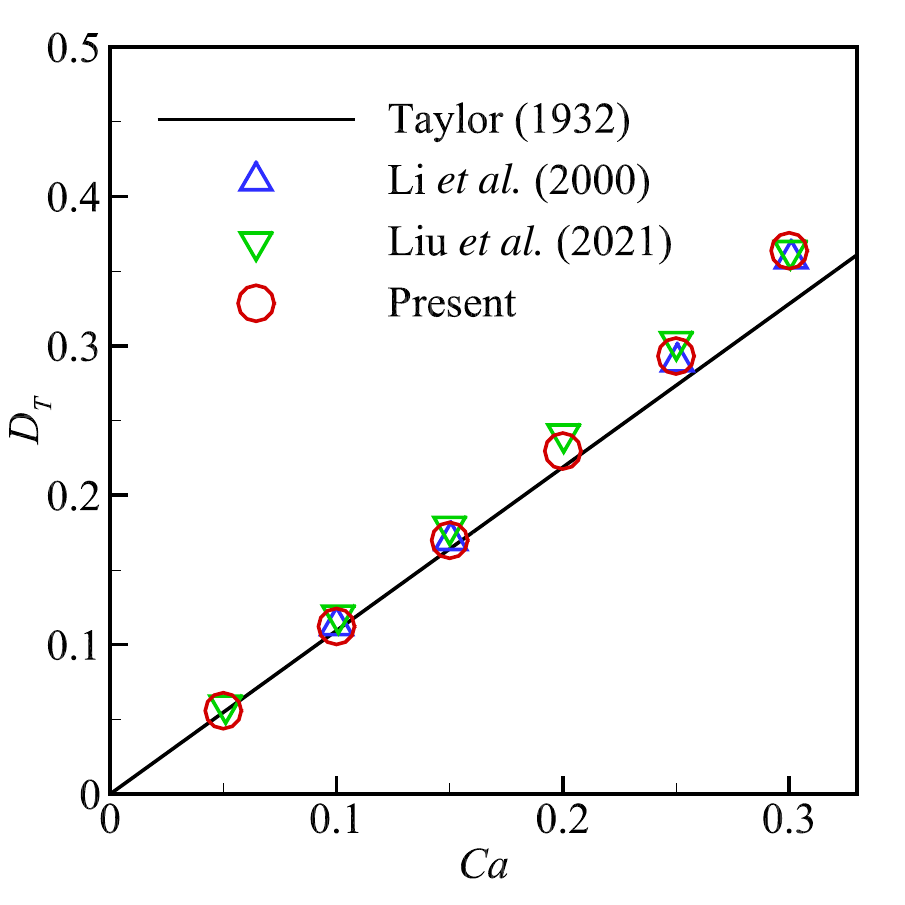}
\put(-180,158){($b$)}
\caption{(a) The interface profile and velocity vectors of the steady-state deformed drop in the $x-z$ plane at $Ca=0.3$ and $Re=0.0625$. The interface is represented by three contours: $C=0.01$, $0.5$ and $0.99$, to indicate the interface is at equilibrium. The reference velocity is shown in the upper left corner. (b) Comparison of the present results with the Taylor's approximate solution (black line) and the previous numerical results~\cite{Li2000pof, Liu2021jcp} in terms of the Taylor deformation parameter $D_T$ at various $Ca$ at $Re=0.0625$.}
\label{case5.ShearFlow(Ca)}
\end{figure}

\begin{figure}[!t]
\centering
\includegraphics[height=2.5cm]{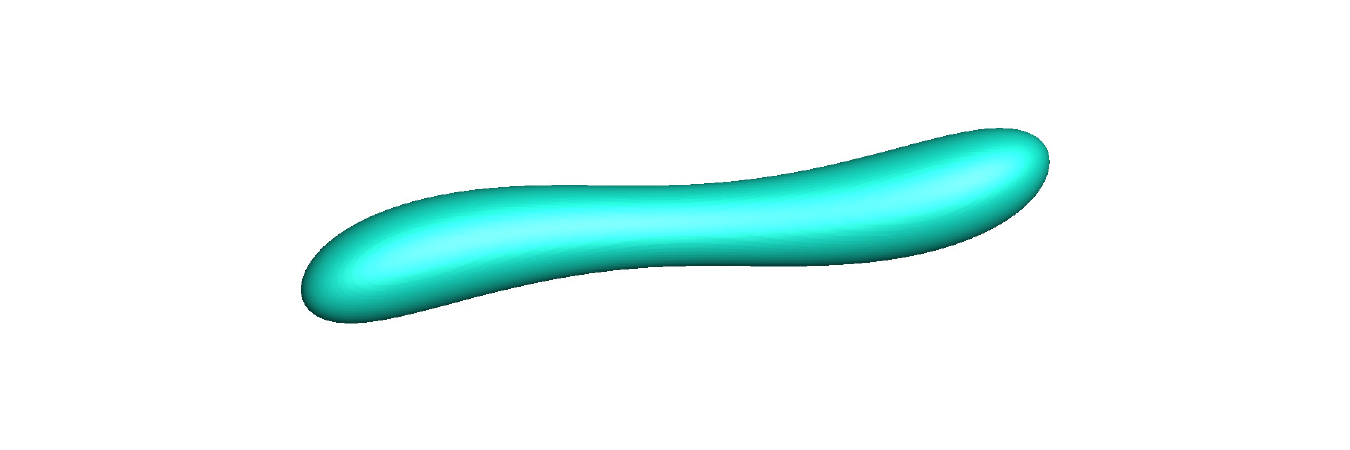}
\put(-200,60){($a$)}\hspace{1cm}
\includegraphics[height=2.5cm]{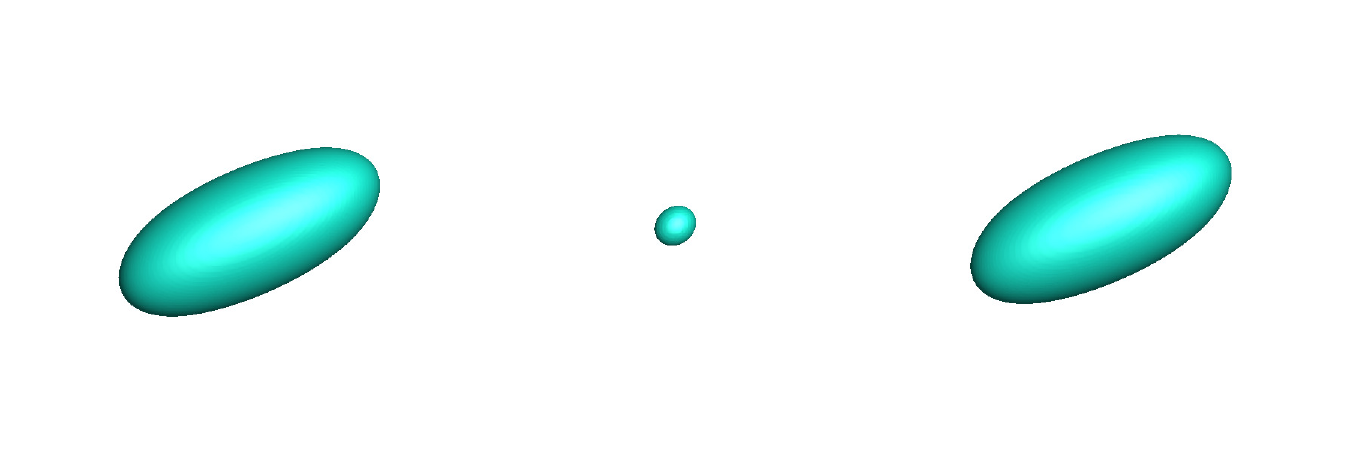}
\put(-200,60){($b$)}
\caption{Snapshots of drop in shear flow at $Ca=0.35$ and $Re=1$: ($a$) $t=25$ and ($b$) $t=50$.}
\label{case5.ShearFlow(BreakUp)}
\end{figure}

Figure~\ref{case5.ShearFlow(Ca)}(a) illustrates the shape and velocity vectors of the steady-state deformed drop in the symmetric $x-z$ plane ($y=0$) at $Ca=0.3$ and $Re=0.0625$. The numerical results of the interface profile in terms of the contours of $C=0.01$, $0.5$ and $0.99$ indicate that the interface maintains a more or less uniform thickness, despite the significant shear flow experienced by the drop. Fig.~\ref{case5.ShearFlow(Ca)}(b) provides the variation of $D_T$ as a function of $Ca$ at $Re=0.0625$, along with Taylor's theoretical prediction and previous results~\cite{Li2000pof, Liu2021jcp}. It is evident that a good agreement with Taylor's linear theory is achieved for $Ca\le 0.15$. The drop deformation at $Ca>0.15$ deviates from Taylor’s prediction, primarily because of the dominance of nonlinear deformation at relatively large $Ca$. These observations are consistent with the other numerical studies~\cite{Li2000pof, Liu2021jcp}. We also simulate the case at $Re=1$ and $Ca=0.35$, and the results at $t=25$ and $50$ are presented in Fig.~\ref{case5.ShearFlow(BreakUp)}. The drop is elongated by the shear flow, then breaks and produces a tiny satellite droplet between two relatively large ones. These phenomena agree well with the numerical results reported by Adami
et al.~\cite{Adami2010jcp}.

\section{Conclusion}
In this paper, we present an improved phase field model for interface capturing in the simulation of two-phase incompressible flows. This model incorporates a second-order diffusion term and a nonlinear diffusion coefficient, which assesses the distortion of the interface profile arising from a non-uniform flow field. To obtain a universal expression of mobility, we make the corresponding scale analysis so that the model can asymptotically approach the sharp interface limit as the interface thickness approaches zero. Furthermore, we propose a general form of smoothed Dirac delta functions that can be adjusted in the thickness of the tension layer while strictly ensuring that its integral equals one, even when the interface profile is not in equilibrium. In addition, we theoretically demonstrate that the spontaneous shrinkage of under-resolved interface structures does not occur in the present model. We systematically validate our method through a series of two-phase flows driven by different flow mechanisms, such as surface tension, buoyancy, Marangoni effect and shear force. The numerical results agree well with experimental data and/or theoretical predictions. Given the wide range of the two-phase flows considered, the recommended range of the dimensionless mobility ($1/Pe\sim 100 Cn^2-333 Cn^2$) can be considered to be universal. Also, the surface tension calculations with the proposed $\delta$ functions have been shown to be accurate and robust.

\section*{Acknowledgments}
We are grateful for the support of the National Natural Science Foundation of China (grant nos. 12302342, 12241204, 12293000, 12293002, 12388101, 12472259) and the Strategic Priority Research Program of the Chinese Academy of Sciences (grant no. XDB0500301).

\bibliographystyle{elsarticle-num} 
\bibliography{References.bib}
\end{document}